\documentclass[onecolumn,12pt]{IEEEtran}
\usepackage[mathscr]{eucal}
\usepackage{ifpdf}
\usepackage{cite}
\usepackage{graphicx}
\usepackage[cmex10]{amsmath}
\usepackage{amssymb}
\usepackage{algorithmic}
\usepackage{units}
\usepackage{setspace}
\usepackage{algorithm}
\usepackage{array}
\usepackage[tight,footnotesize]{subfigure}
\usepackage{amsthm}
\usepackage{multirow}
\usepackage{enumerate}
\usepackage{color}
\newcounter{assume}

\newtheorem{theorem}{Theorem}
\newtheorem{lemma}{Lemma}

\newtheorem{assumption}[assume]{Assumption}

\newtheorem{definition}{Definition}

\begin{document}
\title{Distributed User Association in Energy Harvesting Small Cell Networks: A Competitive Market Model with Uncertainty}
%
\author{
\IEEEauthorblockN{Setareh Maghsudi and Ekram Hossain, \textit{Fellow, IEEE}\\}
\thanks{The authors are with the Department of Electrical and Computer Engineering, University of Manitoba, Winnipeg, MB, Canada (e-mails: \{setareh.maghsudi, ekram.hossain\}@umanitoba.ca). 
}
}
\maketitle
\begin{abstract}
We consider a distributed user association problem in the 
downlink of a small cell network, where small cells obtain the required energy for providing 
wireless services to users through ambient energy harvesting. Since energy 
harvesting is opportunistic in nature, the amount of harvested energy is a random variable, 
without a priori known characteristics. We model the network as a competitive market with 
uncertainty, where self-interested small cells, modeled as consumers, are willing to maximize 
their utility scores by selecting users, represented by commodities. The utility scores of 
small cells depend on the amount of harvested energy, formulated as natures' state. Under 
this model, the problem is to assign users to small cells, so that the aggregate network 
utility is maximized. The solution is the general equilibrium under uncertainty, also called 
Arrow-Debreu equilibrium. We show that in our setting, such equilibrium not only exists, but 
also is unique and is Pareto optimal in the sense of expected aggregate network utility. We 
use the Walras' tatonnement process with some modifications in order to implement the equilibrium 
efficiently.
\end{abstract}
\begin{keywords}
Small cell networks, energy harvesting, uncertainty, user association,  exchange economy,  Arrow-Debreu equilibrium,  
 Walras' tatonnement process. 
\end{keywords}
\section{Introduction}
\label{sec:Introduction}
Due to the ever-increasing need for mobile services, we expect a massive growth in demand 
for wireless data delivery in the years to come. Facing an influx in data traffic, 5G networks 
are foreseen to alleviate this problem by deploying dense small cells to underlay the legacy 
macro cellular networks. This takes advantage from low-power and short-range base stations 
that operate (preferably) using the same radio spectrum as the macro base stations (MBSs) 
and offload macro cell traffic \cite{Hossain14:ETG}, \cite{Fodor12:DAN}. However, deployment 
of small cell networks poses a variety of new challenges to system designers. Examples include 
synchronization \cite{Zou15:NSDS}, resource allocation \cite{Amin15:APG}, interference 
mitigation \cite{Zhang15:ROIM}, handover management \cite{Zhang15:CIM}, and user association. 

In this paper we address the problem of user association in a small cell network where the 
small cells are powered by the energy harvested from the ambient environment. The association 
to the MBS is not considered here. The motivations for the users to associate to the small 
cell base stations (SBSs) include smaller transmission power and the potential of better 
link quality and hence better coverage and capacity. Obviously, the users which are unable 
to associate to any of the SBSs, may associate with the MBS. As user association to the base 
stations is the focus of this paper, in the following we discuss some important related research 
works in more detail. 

In \cite{Namvar14:ACM}, the authors propose a context-aware user-cell 
association approach for small cell networks that exploits the information about the velocity 
and trajectory of users. While taking the quality of service (QoS) requirements into account, 
matching theory is used to design a novel algorithm to solve the user association problem. 
Similarly, in \cite{Semiari14:MTP}, matching theory is applied to solve the user association 
problem in dense small cell networks. Reference \cite{Saad14:CAG} formulates the uplink user 
association as a college admission game and proposes an algorithm based on coalitional games 
to solve the problem. Joint user association and resource allocation is investigated in 
\cite{Chen15:JUA}, where a belief propagation algorithm is proposed for joint user association, 
sub-channel allocation, and power control. Energy-efficient and traffic-aware user association 
are studied in \cite{Mesodiakaki14:EEU} and \cite{Elbassiouny15:TAU}, correspondingly. Both 
studies show the potential of exploiting the available context-aware information, for example, 
users' measurements and requirements, as well as knowledge of the network, to associate the 
users in an energy- and spectrum-efficient way. A cross-layer framework for user association 
control in wireless networks is investigated in \cite{Athanasiou09:ACL}. Reference \cite{Ye13:UAL} 
considers the problem of user association for load balancing. User association in conjunction 
with energy harvesting in small cell network is considered in \cite{Sakr15:AMT}. Therein, stochastic 
geometry is used to develop a modeling framework for $K$-tier uplink cellular networks with RF 
energy harvesting from the concurrent cellular transmissions. 
\subsection{Motivation and Contribution}
\label{subsec:Contribution}
In a vast majority of previous research works, the proposed 
user association scheme is centralized or only partially distributed, which 
necessitates the availability of global channel state information (CSI) at 
a central node, resulting in high computational cost and/or overhead. Therefore, 
it becomes imperative to develop distributed user association schemes that 
are able to cope with information shortage. Furthermore, in order to maintain 
low cost, small cells need to be self-organizing and self-healing. Specifically, 
the required energy of small cells is desired to be harvested locally from the 
ambient environment \cite{Sudeva11:EHS}, rather than being provided by using a fixed power 
supply (which might require frequent recharging), or through energy transfer from 
a power beacon (which might result in transfer costs and energy wastes). This 
sort of energy-independency is in particular feasible in small cell network, 
since small cells normally provide limited services to a small number of users; 
that is, the energy obtained through energy harvesting might suffice to satisfy 
users' requirements. However, since energy harvesting is opportunistic in general, 
uncertainty is a natural attribute of the amount of residual energy in small 
cells. In the presence of uncertainty, distributed user association becomes 
even more challenging, since assignment is performed \textit{before} any 
information regarding the amount of energy in each small cell is disclosed. 
Moreover, it should be mentioned that, in a large body of previous literature, 
the proposed user association method is designated for a specific energy 
harvesting model, for example, random Poisson process \cite{Song14:TUA} 
or Bernoulli energy arrival \cite{Yu15:EHP}. Furthermore, the proposed 
approach depends highly on a specific performance metric, so that by a 
small change in the utility function, the approach is not applicable anymore. 

In the above context, we develop a new theoretical framework 
to analyze the user assignment problem in small cell networks, while taking the 
aforementioned issues into account. We assume that each SBS has the statistical 
CSI of all users with respect to itself, i.e., the local CSI; nevertheless, in 
contrast to centralized approaches, no central controller is in possession of 
global CSI. Moreover, the energy of each small cell is harvested on-site through 
its own energy harvesting units. In order to cope with the hidden uncertainty 
raised by allowing opportunistic energy harvesting, we model the network as a 
competitive market with uncertainty. More precisely, small cells are represented 
by consumers, which selfishly aim at maximizing their utility scores by selecting 
users, despite being uncertain about the amount of harvested energy. Under this 
model, we show the existence of equilibrium and describe its characteristics, 
including optimality and uniqueness. Moreover, based on the well-known Walras' 
tatonnement process, we develop an approach to implement equilibrium efficiently. 
In sharp contrast to previous works, our solution is generic in the 
sense that it does not rely on a specific energy harvesting model, and can be used 
in conjunction with a variety of performance metrics. Moreover, apart from user association, 
our proposed model is applicable to a variety of resource allocation problems under 
uncertainty. An instance is the channel selection problem in cognitive radio networks 
for secondary users, where the statistics of channel availability is unknown a priori. 
In Table \ref{Tb:Comp}, we compare some user association methods in heterogeneous 
networks in conjunction with energy harvesting.\footnote{In Table \ref{Tb:Comp}, uncertainty 
is considered with respect to energy harvesting.}  
%
\begin{table*}[ht]
\caption{User Association in Conjunction with Energy Harvesting in Heterogeneous 
Networks}
\label{Tb:Comp}
\begin{center}
  \small
  \begin{tabular}{|c|c|c|c|c|c|c|}
  \hline
                     &Approach                   &Metric              &Uncertainty&Information&Concentration&Specific Energy Model \\ \hline 
\cite{Liu14:AUA}     &Gradient Descent           &User-Acceptance Rate&   No      &Global CSI &Centralized  &     No               \\ \hline
\cite{Yu15:EHP}      &Stochastic Geometry        &Traffic Offloading  &   No      &Global CSI &Centralized  &     Yes              \\ \hline
\cite{Song14:TUA}    &Stochastic Geometry        &Aggregate Throughput&   No      &Global CSI &Centralized  &     Yes              \\ \hline
\cite{Xu15:DUA}      &Convex Optimization        &Delay-Load Trade-off&   No      &Local CSI  &Distributed  &     Yes              \\ \hline
\cite{Liu14:DDE}     &Convex Optimization        &Delay-Energy Balance&   No      &Global CSI &Centralized  &     Yes              \\ \hline
\cite{Rubio14:UAL}   &Convex Optimization        &Energy Efficiency   &   No      &Global CSI &Centralized  &     Yes              \\ \hline
Our Work             &Exchange Economy           &Aggregate Throughput&   Yes     &Local CSI  &Distributed  &     No               \\ \hline  
  \end{tabular}
 \end{center} 
\end{table*}  
\subsection{Paper Organization}
\label{subsec:Organization}
The rest of the paper is organized as follows. In Section \ref{sec:System}, we describe 
the system model and formulate the user association problem. In Section \ref{sec:Wahlras}, 
we briefly introduce competitive market model and exchange economy under uncertainty, by 
providing some basic definitions and results. We also describe the Walras' tatonnement 
procedure. In Section \ref{sec:CMModel}, we model the user association problem as an 
exchange economy under uncertainty, and we characterize the equilibrium. We also show 
that in an exchange economy with indivisible goods, the demand calculation can be formulated 
and solved as a static knapsack problem. Section \ref{sec:Num} includes numerical results. 
Section \ref{sec:con} summarizes the paper and adds some concluding remarks.
\subsection{Notation}
\label{subsec:Notation}
Throughout the paper we denote a set and its cardinality by a unique letter, and 
distinguish them by using calligraphic and italic fonts, such as $\mathcal{A}$ and 
$A$, respectively. Matrices are shown by bold upper case letters, for instance 
$\mathbf{A}$. Moreover, $\mathbf{A}_{l}$ denotes the $l$-th row of matrix $\mathbf{A}$. 
$[\mathbf{A}]_{l,m}$ stands for the element of matrix $\mathbf{A}$ located at $l$-th 
row and $m$-th column. Unit vectors are shown by bold lower case letters, for example, 
$\textbf{a}$. The Hadamard (element-wise) product of two matrices $\mathbf{A}$ and 
$\mathbf{B}$ is shown as $\mathbf{A}{\circ}\mathbf{B}$. Furthermore, by $\mathbf{A}=
\mathbf{B^{^{\circ\frac{1}{2}}}}$ we mean that matrix $\mathbf{B}$ is the element-wise 
square of matrix $\mathbf{A}$.
\section{System Model and Problem Formulation}
\label{sec:System}
We consider a small cell network consisting of an MBS, a set $\mathcal{M}=\{1,...,M\}$ 
of users and a set $\mathcal{N}=\{1,...,N\}$ of small cells. Every small cell $n \in 
\mathcal{N}$ provides wireless services to a set of users, $\mathcal{M}_{n} \subseteq 
\mathcal{M}$. Define the assignment matrix $\mathbf{X}=\left[x_{nm}\right]_{N \times M}$. 
Traditionally, each user can associate to a single SBS; that is, 
\begin{equation}
\label{eq:ConsF}
\begin{matrix}
x_{nm}\in \left \{0,1\right \}, &\forall ~m \in \mathcal{M},n \in \mathcal{N}
\end{matrix},
\end{equation}
and, 
\begin{equation}
\label{eq:ConsS}
\begin{matrix}
\sum_{n \in \mathcal{N}}x_{nm}= 1, &\forall ~m \in \mathcal{M}
\end{matrix}.
\end{equation}
Multiple simultaneous associations, however, would enhance the system throughput 
and reduce the outage ratio, particularly for cell edge users \cite{Hossain14:ETG}. 
Therefore, in this paper, we develop a framework that can also be applied to the 
continuous case, where, from every user $m \in \mathcal{M}$, a fraction $x_{nm} \in 
\left [0,1 \right]$ is assigned to small cell $n \in \mathcal{N}$; that is,
\begin{equation}
\label{eq:ConsTC}
\begin{matrix}
x_{nm}\in \left [0,1 \right], &\forall ~m \in \mathcal{M},n \in \mathcal{N}
\end{matrix},
\end{equation}
where also (\ref{eq:ConsS}) holds.
The fraction $x_{nm}$ can be interpreted as some sort of handover: Small cell 
$n$ provides $x_{nm}$ fraction of the required spectrum resources for user $m$, 
or supports it for $x_{nm}$ fraction of the transmission time. As mentioned 
before, by $\mathbf{X}_{n}$, we denote the $n$-th row of matrix $\mathbf{X}$, 
which provides information about the users assigned to small cell $n \in \mathcal{N}$. 
Clearly, for the discrete and continuous cases, we have $\mathbf{X}_{n} \subset 
\{0,1\}^{M}$ and $\mathbf{X}_{n} \subset \left [0,1 \right]^{M}$, respectively. 
In our work, we distinguish those results and algorithms that cannot be directly 
applied to the continuous model, and provide substitute solutions. 

We assume that energy harvesting is independent across small cells. In other words, 
each small cell $n \in \mathcal{N}$ has its own energy harvesting units, so that by 
utilizing these units, its required energy is locally obtained from the ambient 
environment. The energy is then spent by the small cell to serve its assigned users. 
We assume that the network operates in two consecutive time intervals. In the first 
interval, user association and scheduling decisions are made; that is done while SBSs 
are harvesting the energy, but before any information about the amount of harvested 
energy is disclosed. Transmissions are performed in the second interval, after each 
SBS knows the exact amount of harvested energy. In general and regardless of the 
specific method, energy harvesting is opportunistic; hence, for every small cell, the 
amount of harvested (or residual) energy is random in nature. Therefore, we model the 
residual energy as a random variable. We do not make any assumption on the probability 
density function of this random variable. For each small cell $n \in \mathcal{N}$, we 
refer to every possible level of residual energy as one state, which belongs to a finite 
set of integer values $\mathcal{S}_{n}$. The state space of the environment can be then 
written as 
\begin{equation}
\label{eq:stateModel}
\mathcal{S}=\bigotimes_{n=1}^{N}\mathcal{S}_{n},
\end{equation}
where $\bigotimes$ is the Cartesian product. From (\ref{eq:stateModel}), at each time, 
the nature's state is defined as the collection of residual energy levels of all small 
cells. 

We assume that each small cell is provided with sufficient spectrum 
resources to guarantee orthogonal transmission to its assigned users; that is, inside 
every small cell, transmissions are corrupted only by zero-mean additive white Gaussian 
noise (AWGN) with variance $N_{0}$. For each small cell $n \in \mathcal{N}$, the intercell 
interference experienced by every user $m \in \mathcal{M}_{n}$, denoted by $I_{nm} \geq 0$, 
is regarded as noise and is assumed to be fixed and known. The average channel gain between 
user $m \in \mathcal{M}$ and small cell $n \in \mathcal{N}$ (including Rayleigh fading and 
path loss) is denoted by $h_{nm}$. We assume that each SBS $n \in \mathcal{N}$ is provided 
with statistical CSI of every link $n \to m$ for all $m\in \mathcal{M}$. We focus on downlink 
transmission. The model can be used in combination with various power allocation and/or 
transmission scenarios in each cell. As few examples, we consider the following transmission 
scenarios.\footnote{These scenarios are only mentioned as examples, and the application 
of the proposed model and solution is not limited to these scenarios.}
\begin{itemize}
\item Fixed-power transmission: Every SBS transmits to each of its 
assigned users with some fixed power $P_{n}$. Since the total available transmission 
power is a monotone increasing function of the bounded harvested energy, every SBS 
is able to serve only a limited number of users. Let $P^{(s)}_{n} \in \mathcal{R}_{+}
-\left \{\infty \right \}$ be the total available power at small cell $n \in \mathcal{N}$. 
Thus, at most $\alpha^{(s)}_{n}=\left \lfloor \frac{P^{(s)}_{n}}{P_{n}} \right \rfloor$ 
users can be served. If $\alpha^{(s)}_{n}<M_{n}$, then a subset of users, say $\mathcal{M}'_{n} 
\subset \mathcal{M}_{n}$ has to be selected to be served, where $M'_{n} \leq \alpha^{(s)}_{n}$. 
For each $m \in \mathcal{M}_{n}$, the achievable transmission rate is given by 
\begin{equation}
\label{eq:TransRateT}
r_{nm}(h_{nm})=\log \left(1+ \frac{P_{n}h_{nm}}{N_{0}+I_{nm}} \right).
\end{equation}
To define $\mathcal{M}'_{n}$, each small cell orders the assigned users based on achievable 
transmission rate. Let $\mathcal{M}_{n,\max}^{(s)} \subseteq \mathcal{M}_{n}$ denote the set 
of $\alpha^{(s)}_{n}$ users that yield the largest transmission rates; that is, the first 
$\alpha^{(s)}_{n}$ users in the descending ordering of achievable transmission rates are 
collected in the set $\mathcal{M}_{n,\max}^{(s)}$. The utility of small cell $n$ from user $m$ 
is then defined as
\begin{equation}
\label{eq:Utility}
u_{nm}^{(s)}=\begin{cases}
r_{nm} & m \in \mathcal{M}_{n,\max}^{(s)} \\ 
0 & \text{o.w.}
\end{cases}.
\end{equation}
That is, only the best $\alpha^{(s)}_{n}$ users are served, which thus yields 
benefit. 

\item Full-power transmission: Each SBS can broadcast the data with the entire 
available power to all its assigned users, in case every user in a small cell is interested in 
receiving all information. Such scenario arises often in multimedia transmission, online gaming, 
and file downloading. In this case, at every state $s$, the utility of SBS $n$ from every user 
$m \in \mathcal{M}_{n}$ is given by (\ref{eq:TransRateT}) after replacing $P_{n}$ with $P^{(s)}_{n}$.

\item Equal-power transmission: Every SBS transmits to its entire assigned users 
with equal power; that is, the available power is shared by all users equally. Then, at each state 
$s$, the utility is given by (\ref{eq:TransRateT}) with $P_{n}=P^{(s)}_{n}/M_{n}$.
\end{itemize}
Regardless of the transmission model, for every small cell $n \in \mathcal{N}$, we define the 
utility function $u_{n}: \{0,1\}^{M} \to \mathcal{R}_{+}-\left \{\infty \right \}$ (or $\left[0,1\right]^{M} 
\to \mathcal{R}_{+}-\left \{\infty \right \}$, for the continuous case) as the aggregate data 
rate experienced by the users assigned to that specific small cell. Formally,
\begin{equation}
\label{eq:UtiSmal}
u_{n}^{(s)}(\mathbf{X}_{n})=\sum_{m \in \mathcal{M}}u_{nm}^{(s)}[\mathbf{X}]_{n,m},
\end{equation}
where $[\mathbf{X}]_{n,m}$ stands for the element of matrix $\mathbf{X}$ located at $n$-th 
row and $m$-th column.  

As described before, user association is performed before energy harvesting; that is, at 
the time of user association, the amount of energy in each small cell, or, in other words, 
the nature's state, is unknown. Under this uncertainty, the problem is to assign users to 
small cells in a distributed manner. By (\ref{eq:UtiSmal}), for every small 
cell, the utility is state-dependent. Thus, the assignment matrix might be state-dependent 
as well. For any state $s \in \mathcal{S}$, we define the assignment matrix as $\mathbf{X}^{(s)}
=\left [x_{nm}^{(s)}\right]_{N \times M}$. Let $\mathcal{X}$ be the set of all possible assignment 
matrices. Ideally, desired is to find $x_{nm}^{(s)}$ for all $m \in \mathcal{M}$ and $n \in \mathcal{N}$, 
so that at every state $s \in \mathcal{S}$, $\mathbf{X}^{(s)}$ is a solution of the following 
optimization problem:
\begin{equation}
\label{eq:OptPr}
\underset{\mathbf{X}^{(s)} \in \mathcal{X}}{\textup{maximize}}~\sum_{n \in \mathcal{N}} 
u_{n}^{(s)}\left(\mathbf{X}_{n}^{(s)}\right),
\end{equation}
where for the discrete problem formulation, the problem is subject to (\ref{eq:ConsF}) 
and (\ref{eq:ConsS}), while (\ref{eq:ConsTC}) and (\ref{eq:ConsS}) are the constraints 
for the continuous case. 

However, since the nature's state is unknown a priori, (\ref{eq:OptPr}) is 
infeasible. Therefore, we opt for a less ambitious goal, described in the following. Assume 
that each small cell $n \in \mathcal{N}$ assigns probability vector $\mathbf{a}_{n}=\left(
a_{n}^{(1)},...,a_{n}^{(S)}\right)$ to the set of states, $\mathcal{S}$. Then the expected 
utility yields 
\begin{equation}
\label{eq:AvNetUt}
\bar{u}_{n}=\sum_{s=1}^{S}a_{n}^{(s)}u_{n}^{(s)}\left(\mathbf{X}_{n}^{(s)} \right).
\end{equation}
We define the goal of user association as follows: For every state $s$, find 
an assignment matrix $\mathbf{X}$ that is a solution of the following optimization problem:
\begin{equation}
\label{eq:OptPrT}
\underset{\mathbf{X}^{(s)} \in \mathcal{X}}{\textup{maximize}}~\sum_{n \in \mathcal{N}} 
\bar{u}_{n}\left(\mathbf{X}_{n}^{(s)} \right),
\end{equation}
subject to (\ref{eq:ConsF}), (\ref{eq:ConsS}) for the discrete problem and (\ref{eq:ConsTC}), 
(\ref{eq:ConsS}) for the continuous case. That is, the goal is defined as to maximize the aggregate 
expected utility of all SBSs.
\section{Exchange Economy with Uncertainty}
\label{sec:Wahlras}
According to the system model described in Section \ref{sec:System}, the problem 
is to assign users to small cells, \textit{before} any information about energy 
harvesting is disclosed, in a way that the expected aggregate network utility is 
maximized. We aim at solving the problem in a distributed manner, by allowing 
every small cell to select users on its own. In doing so, the selfishness of 
small cells should be taken into account: Naturally, every small cell would like 
to select as many users as possible so as to maximize its own reward given by 
(\ref{eq:UtiSmal}); nonetheless, the utility that can be achieved by each small 
cell, or, in other words, the contribution of each small cell to the network 
aggregate utility, is dictated by the amount of harvested energy in that small 
cell. The energy, however, is obtained through energy harvesting rather than a 
fixed power supply, and therefore can be regarded as a random variable whose 
realization is unknown a priori. Therefore, a mechanism is required to prevent 
SBSs from overestimating their energy resources while selecting users.

In this paper, we model the small cell network as a \textit{competitive market}, 
accommodating an \textit{exchange economy with uncertainty}. In this model, there 
is no producer. Consumers, each provided with an initial endowment, trade the 
available commodities while being ambitious to maximize their own utilities, and 
uncertain about the nature's state. Under this model, the solution of problem (\ref{eq:OptPr}) 
is the \textit{general equilibrium under uncertainty}, also called \textit{Arrow-Debreu 
equilibrium}.

In this section, we describe some basic notions of exchange economy and equilibrium 
under uncertainty. In Section \ref{sec:CMModel}, we describe our proposed model and 
analyze the formulated user association problem based on exchange economy.   
\subsection{Exchange Economy}
\label{subsec:Competitive}
An exchange economy $\Omega$ consists of a set of consumers, $n \in \mathcal{N}$, and a set 
of commodities (goods), $m \in \mathcal{M}$. Let $\mathbf{I}=\left(i_{1},...,i_{M} \right)$ 
stand for the endowment of goods, with $i_{m} \in \mathcal{Z}$, $m \in \mathcal{M}$. For each 
consumer $n \in \mathcal{N}$, a \textit{bundle} of commodities is a row matrix $\mathbf{X}_{n} 
\in \mathcal{Z}^{M}$ for indivisible goods and $\mathbf{X}_{n} \in \mathcal{R}_{+}^{M}$ for 
divisible goods. Each consumer is supplied with an initial endowment of commodities gathered 
in a row matrix $\mathbf{Q}_{n} \in \mathcal{Z}^{M}$ or $\textbf{Q}_{n} \in \mathcal{R}_{+}^{M}$ 
for indivisible and divisible goods, respectively, so that $\sum_{n \in \mathcal{N}}\mathbf{Q}_{n}=
\mathbf{I}$. The model captures the idea of exchanging goods, without production, where the 
allocation of a given amount of each commodity implies its final consumption, associated with 
some utility score. A map $u_{n}:\mathcal{Z}^{M} \to \mathcal{R}$ (or $u_{n}: \mathcal{R}_{+}^{M}
\to \mathcal{R}$, for divisible goods) is called a utility function on the commodity set $\mathcal{M}$ 
for every consumer $n \in \mathcal{N}$. Since utilities and initial endowments are the main blocks 
of exchange economy, we define any exchange economy $\Omega$ as $\Omega:\left\{u_{n},\mathbf{Q}_{n}
\right\}_{n \in \mathcal{N}}$. In the following we define the \textit{monotonicity} of utility 
functions.
\begin{definition}[Monotonicity]
\label{de:Mo}
A utility function $u_{n}: \mathcal{Z}^{M} \to \mathcal{R}$ is monotone if for all 
$\mathbf{X}^{'}_{n} \leq \mathbf{X}_{n}, u_{n} (\mathbf{X}^{'}_{n})\leq u_{n} 
\left(\mathbf{X}_{n}\right)$.
\end{definition}

In a deterministic model of exchange economy, all information are provided to all 
individuals a priori, whereas under uncertainty, the nature can have different states, 
modeled as the outcomes of some random variable, unknown a priori. Let $\mathcal{S}$ 
denote the set of all possible states of the nature. We make the following assumption.
\begin{assumption}
\label{ass:stateSpace}
We assume the following on the set of space's states, $\mathcal{S}$.
\begin{enumerate}[(a)]
 \item $\mathcal{S}$ is a finite set.
 \item All elements in $\mathcal{S}$ are mutually exclusive.
 \item $\mathcal{S}$ is exhaustive.
 \item $\mathcal{S}$ is known to all individuals.\footnote{The relaxation 
       of this assumption makes the problem more challenging and is to be 
       investigated in our future work.}
\end{enumerate} 
\end{assumption}
For $S$ states of nature and $M$ commodities, there exist $MS$ \textit{state contingent} 
commodities. In simple words, state contingency means that a commodity $m$ in state $s$ 
is regarded as another commodity, say $m'$, when the state changes to $s'$. Arrow and 
Debreu show that the basic notions and theorems for a deterministic exchange economy with 
$M$ goods, for instance the two Welfare theorems \cite{Colell85:MT}, hold also for an 
exchange economy under uncertainty with $S$ possible space states and $M$ goods, simply 
by building a new exchange economy with $MS$ goods, i.e., state contingent commodities 
\cite{Debreu87:TTV}. In such an economy, we define the \textit{allocation} of a state 
contingent commodity as follows.
\begin{definition}[Allocation]
\label{de:ConCom}
An allocation of state contingent commodity $ms$, denoted by $x_{nm}^{(s)} \in \mathcal{Z}$ 
(or $x_{nm}^{(s)} \in \mathcal{R}_{+}$, for divisible goods), describes the number of units 
(or the amount) of commodity $m \in \mathcal{M}$ that can be consumed by individual $n \in 
\mathcal{N}$, if and only if state $s \in \mathcal{S}$ occurs.
\end{definition}
Therefore, a state contingent allocation vector can be written as $\mathbf{X}_{n}= 
\left(x_{n1}^{(1)},...,x_{nM}^{(1)},...,x_{n1}^{(S)},...,x_{nM}^{(S)} \right)$. 
Throughout the paper, at each state $s \in \mathcal{S}$, we denote the vector of 
commodities allocated to consumer $n \in \mathcal{N}$ as $\mathbf{X}_{n}^{(s)}$. 
In a competitive market with uncertainty, the initial endowment of each consumer 
is in general random and state-dependent. Formally, initial endowments yields 
$\mathbf{Q}_{n} =\left(q_{n1}^{(1)},...,q_{nM}^{(1)},...,q_{n1}^{(S)},...,q_{nM}^{(S)}\right)$. 
Moreover, $\mathbf{Q}_{n}^{(s)}$ is the vector of initial endowments of consumer 
$n \in \mathcal{N}$ at state $s \in \mathcal{S}$. Every individual has to pay a 
price for every unit of each commodity under each state. That is, a price $p_{m}^{(s)}$ 
has to be paid for every unit of a state contingent commodity $ms$, $m \in \mathcal{M}$ 
and $s \in \mathcal{S}$. Hence we denote the price vector by $\mathbf{P}=\left(p_{1}^{(1)},
...,p_{M}^{(1)},...,p_{1}^{(S)},...,p_{M}^{(S)}\right) \in \mathcal{R}_{+}^{MS}$. 
Also, at each state $s$, the price vector is denoted by $\mathbf{P}^{(s)} \in 
\mathcal{R}_{+}^{M}$. 

Now assume that consumer $n \in \mathcal{N}$ assigns probability vector $\mathbf{a}_{n}
=\left( a_{n}^{(1)},...,a_{n}^{(S)} \right)$ to the set of states, $\mathcal{S}$. Consumers 
do not have to agree on a specific probability distribution. So, the expected utility 
is given by (\ref{eq:AvNetUt}). Then, consumers' preferences, denoted by $\succeq_{n}$ 
on $\mathcal{R}_{+}^{MS}$, $n \in \mathcal{N}$, are state-dependent and can be represented 
as
\begin{equation}
\label{eq:Preferen}
\mathbf{X'}_{n}^{(s)}\succeq_{n} \mathbf{X}_{n}^{(s)} \Leftrightarrow \bar{u}_{n}^{(s)} 
\left(\mathbf{X'}_{n}^{(s)}\right)\geq \bar{u}_{n}^{(s)}\left(\mathbf{X}_{n}^{(s)} \right).
\end{equation}
Let the utility of each consumer $n \in \mathcal{N}$ from commodity $m \in \mathcal{M}$ be 
denoted by $u_{nm}$. Then, at state $s$, its net benefit yields
\begin{equation}
\label{eq:UtiSmallT}
v_{n}^{(s)}\left(\mathbf{X}_{n}^{(s)},\mathbf{P}^{(s)}\right)=\sum_{m \in \mathcal{M}}u_{nm}
\left[\mathbf{X}_{n}^{(s)}\right]_{1,m}- \sum_{m \in \mathcal{M}}\left[\mathbf{P}^{(s)}\right]_{1,m}.
\end{equation}
At each state $s \in \mathcal{S}$, given the prices of commodities, the \textit{demand correspondence} 
of a consumer $n \in \mathcal{N}$ is defined as  
\begin{equation}
\label{eq:demCor}
\mathcal{D}_{n}^{(s)}\left(\mathbf{P}^{(s)}\right):=\left \{ \mathbf{X}_{n}^{(s)}|
v_{n}^{(s)}\left(\mathbf{X}_{n}^{(s)},\mathbf{P}^{(s)} \right)\geq v_{n}^{(s)} 
\left(\mathbf{X'}_{n}^{(s)},\mathbf{P}^{(s)}\right)\right\}, 
\end{equation}
where $\mathbf{X}_{n}^{(s)} \in \mathcal{Z}^{M}$ and $\mathbf{X}_{n}^{(s)} \in \mathcal{R}_{+}^{M}$ 
for indivisible and divisible commodities, respectively. Thus the final consumption depends on the 
state of nature; this means that same quantities of a given commodity might result in different 
utility scores in different states. We now define \textit{gross substitutes}, which is an important 
concept in exchange economy.
\begin{definition}[Gross Substitutes]
\label{de:GS}
Let $\mathcal{D}_{n}\left(\mathbf{P}\right)$ be the demand correspondence for consumer 
$n$, given price vector $\mathbf{P}$, defined by (\ref{eq:demCor}). A utility function 
$u_{n}: \mathcal{Z}^{M} \to \mathcal{R}$ (or $u_{n}:\mathcal{R}^{M}_{+} \to \mathcal{R}_{+}$) 
satisfies the gross substitutes condition if for any two price vectors $\mathbf{P}$ and 
$\mathbf{P}'$ where $\mathbf{P}' \geq \mathbf{P}$, and for any $\mathbf{X}_{n} \subset 
\mathcal{D}_{n}\left(\mathbf{P}\right)$, there exists $\mathbf{X}'_{n} \subset \mathcal{D}_{n}
\left(\mathbf{P}'\right)$ such that for all $m \in \mathcal{M}$, $x_{nm}=x'_{nm}$ if $p_{m}
=p'_{m}$.
\end{definition}
In words, \textit{gross substitutes} means that if a consumer demands a bundle of commodities 
and the prices of some good increase, it would still demand the goods in that bundle whose 
prices did not change.

Next we state some conventional assumptions on the utility function.
\begin{assumption}
\label{ass:Uti}
For all $n \in \mathcal{N}$ and $u_{n}:\mathcal{R}_{+}^{M}\to \mathcal{R}$,
\begin{enumerate}[(a)]
 \item $u_{n}$ is continuous, 
 \item $u_{n}$ is increasing, 
 \item $u_{n}$ is concave, and
 \item $\mathbf{Q}_{n}>0$.
\end{enumerate} 
\end{assumption}
\begin{assumption}
\label{ass:UtiId}
For all $n \in \mathcal{N}$ and $u_{n}: \mathcal{Z}^{M} \to 
\mathcal{R}$,  
\begin{enumerate}[(a)]
 \item $u_{n}$ is monotone, and
 \item it satisfies the gross substitutes condition.      
\end{enumerate} 
\end{assumption}

We gather all individual demands $\left[\mathbf{X}_{n}\right]_{1\times M}$, $n \in 
\mathcal{N}$, in a matrix $\left[\mathbf{X}\right]_{N\times M}$. Each consumer selects 
its consumption so as to maximize its expected utility; that is, consumer $n \in 
\mathcal{N}$ solves for 
\begin{equation}
\label{eq:UserOpt}
\begin{aligned}
\underset{\mathbf{X}_{n}}{\textup{maximize}} & ~~ \bar{u}_{n}(\mathbf{X}_{n}) \\ 
 \textup{s.t.}~~~~&\mathbf{P} \cdot \mathbf{X}_{n}\leq \mathbf{P} \cdot \mathbf{Q}_{n}
\end{aligned},
\end{equation}
where $\mathbf{P}.\mathbf{X}_{n}\leq \mathbf{P}.\mathbf{Q}_{n}$ is the budget constraint. In general, 
the budget set writes
\begin{equation*}
\label{eq:BudSet}
\mathcal{B}_{n}(\mathbf{P})=\left \{\mathbf{X}_{n}:\mathbf{P}\cdot \mathbf{X}_{n} \leq \mathbf{P} 
\cdot\mathbf{Q}_{n} \right \}.
\end{equation*}
At some price vector $\mathbf{P}$ and for a good $m \in \mathcal{M}$, the excess demand is defined 
as
\begin{equation}
\label{eq:}
z_{m}(\mathbf{P})= \sum_{n\in \mathcal{N}}d_{nm}(\mathbf{P}, \mathbf{P}.\mathbf{Q}_{n})-
\sum_{n \in \mathcal{N}} q_{nm},
\end{equation}
where $d_{nm}(\cdot)$ is the demand of consumer $n \in \mathcal{N}$ for good $m \in \mathcal{M}$. Now 
we define \textit{Pareto optimal allocation}. 
\begin{definition}[Feasible Allocation]
\label{de:FAllocation}
In an exchange economy $\Omega$, an allocation $\mathbf{X}$ is feasible if $\sum_{n \in \mathcal{N}} 
\mathbf{X}_{n}\leq \sum_{n \in \mathcal{N}} \mathbf{Q}_{n}$. 
\end{definition}
\begin{definition}[Pareto Optimal Allocation]
\label{de:PrOptimal}
A feasible allocation $\mathbf{X}$ is Pareto optimal if there is no other feasible allocation 
$\mathbf{X}'$ such that $\bar{u}_{n}(\mathbf{X'}_{n}) \geq \bar{u}_{n}(\mathbf{X}_{n})$ for all 
$n \in \mathcal{N}$ with strict inequality for some $n \in \mathcal{N}$.
\end{definition}
Before we proceed to the next section, we state two lemmas that we use later to prove some characteristics 
of our proposed solution.
\begin{lemma}[\cite{Gul99:WEGS}]
\label{Lm:GSAC}
Additive concave and separable additive utility functions satisfy the gross substitutes property.
\end{lemma}
\begin{lemma}[\cite{Levin06:GE}]
\label{Lm:ExcDem}
If each individual has a utility function satisfying the gross substitutes condition, then both 
the individual and aggregate excess demand functions satisfy gross substitutes condition as 
well.
\end{lemma}
\subsection{Arrow-Debreu Equilibrium}
\label{subsec:Equi}
The outcome of a deterministic competitive market (exchange economy) is the \textit{general 
equilibrium}, also called \textit{Walrasian equilibrium}. In an uncertain environment, 
however, deterministic equilibrium is insufficient, and extensions should be provided. 
One extension is the \textit{general equilibrium under uncertainty}, also called \textit{Arrow-Debreu 
equilibrium}. In this notion of equilibrium, prices are not state contingent, 
whereas allocations are. In other words, commitments are made before the state is known, at date 
0, while allocations follow after revealing the state, at date 1. The sequence of events in an 
Arrow-Debreu model is described in Algorithm \ref{de:ArDeSe}. 
\begin{algorithm}
\caption{Arrow-Debreu Sequence of Events}
\label{de:ArDeSe}
\small
\begin{algorithmic}[1]
\STATE At $t=0$, nature's state is unknown. Commitments are however made in this stage, i.e., 
       under the uncertainty. This means that, at $t=0$, each agent $n \in \mathcal{N}$ 
      determines its optimal demand and buys $x_{nm}^{(s)}$ units (amount) of each good $m \in 
      \mathcal{M}$ if state $s \in \mathcal{S}$ occurs, and signs a contract in which it commits 
      to pay the price. 
\STATE At $t=1$, the state $s$ is revealed. The contracts under state $s$ are executed, and all 
      other contracts become void; that is, each individual receives the purchased goods under 
      current state and pays the associated price.
\end{algorithmic}
\end{algorithm}

For a competitive market, the \textit{Arrow-Debreu equilibrium} is defined as follows \cite{Debreu87:TTV}.
\begin{definition}[Arrow-Debreu Equilibrium]
\label{de:ArDeEq}
An allocation matrix $\mathbf{X}$, together with a price vector $\mathbf{P}$ are an Arrow-Debreu 
equilibrium if 
\begin{enumerate}
 \item $\forall n \in \mathcal{N}$, $\mathbf{X}_{n}$ maximizes $\bar{u}_{n}$ on $\mathcal{B}_{n}$;
 \item Market clears, that is, for all $m \in \mathcal{M}$
       \begin{equation}
       \label{eq:FeasThrMC}
       \sum_{n \in \mathcal{N}} x_{nm} = \sum_{n \in \mathcal{N}} q_{nm}.
       \end{equation}  
\end{enumerate}
\end{definition}
Now we are in a position to describe some results regarding the existence 
and optimality of Arrow-Debreu equilibrium. It should be mentioned that the 
following results are originally developed for the outcome of a deterministic 
competitive market model, i.e., \textit{Walrasian} equilibrium \cite{Gul99:WEGS}. 
Nonetheless, in \cite{Debreu87:TTV}, it is shown that these results hold also 
for competitive market with uncertainty, with its outcome being \textit{Arrow-Debreu} 
equilibrium. 
\begin{theorem}[Existence of Arrow-Debreu Equilibrium]
\label{Th:Ex}
Arrow-Debreu equilibrium $\left(\mathbf{X}, \mathbf{P} \right)$ exists for
\begin{enumerate}[(a)]
\item an economy with divisible goods that satisfies Assumption (\ref{ass:Uti}) 
      \cite{Colell85:MT}, and 
\item an economy with indivisible goods that satisfies Assumption (\ref{ass:UtiId}) 
      \cite{Gul99:WEGS}.
\end{enumerate}
\end{theorem}
\begin{theorem}[Optimality of Arrow-Debreu Equilibrium]
\label{Th:Opt}
Let $\left(\mathbf{X},\mathbf{P} \right)$ be an Arrow-Debreu equilibrium. Then,
\begin{enumerate}[(a)]
\item the allocation $\mathbf{X}$ is Pareto optimal for an economy $\Omega$ with divisible goods if 
      Assumption \ref{ass:Uti}-b holds \cite{Gul99:WEGS}.
\item the allocation $\mathbf{X}$ is Pareto optimal for an economy $\Omega$ with indivisible goods 
      if Assumption \ref{ass:UtiId}-a holds \cite{Colell85:MT}; 

\end{enumerate}
\end{theorem}
\begin{theorem}[Uniqueness of Arrow-Debreu Equilibrium \cite{Levin06:GE}]
\label{Th:Uniq}
For an economy $\Omega$ with divisible goods, if the aggregate excess demand function 
$z(\cdot)$ satisfies the gross substitutes condition, then the economy has at most one 
Arrow-Debreu equilibrium, i.e., $z\left(\mathbf{P} \right)=0$ has at most one (normalized) 
solution. 
\end{theorem}

In the next section we describe a method to achieve equilibrium in an exchange economy.
\subsection{Walras' Tatonnement Process}
\label{subsec:Auction}
In previous sections, we discussed an exchange economy under uncertainty, and 
described the notion of equilibrium in such setting. Still, it is important to 
describe how the equilibrium prices are reached, or, in other words, how equilibrium 
is implemented. To this end, Walras developed a price adjustment process, namely 
\textit{Walras' tatonnement}, known also as \textit{Walrasian auction}. The process 
requires a coordinator, called \textit{Walrasian auctioneer}, which, at each round, 
announces the prices, starting at some random initial point. Afterwards, consumers 
disclose their demands at those prices, so that the auctioneer can adjust prices to 
demands. The process continues until market clears; that is, when a set of prices 
yields a demand equal to supply. At this point, prices and demands are final, and the 
auction process terminates, i.e., trade occurs \cite{Uzawa60:WTT}. Let $z \left(\mathbf{P}\right)$ 
be the excess demand given price vector $\mathbf{P}$. For divisible goods, the price 
adjustment rule for the auctioneer is \cite{Levin06:GE}
\begin{equation}
\label{eq:AuctionO}
\mathbf{P}(t+1)= \mathbf{P}(t)+ \alpha z \left(\mathbf{P}(t)\right),
\end{equation}
for a sufficiently small $\alpha>0$. Clearly the only stationary points of this process are 
prices $\mathbf{P}$ at which $z(\mathbf{P})=0$, i.e., equilibrium prices \cite{Levin06:GE}. 
For indivisible goods, the price adjustment rule yields \cite{Mochaourab15:DCA}
\begin{equation}
\label{eq:AuctionT}
\mathbf{P}(t+1)= \mathbf{P}(t)+ \alpha \Delta \left(\mathbf{P}(t)\right),
\end{equation}
where $\Delta=\left(\delta_{1},...,\delta_{M}\right)$, and $\delta_{m}=1$ if $m \in \mathcal{M}$ 
is in excess demand and zero otherwise. The procedure is summarized in Algorithm \ref{Alg:Auction}. 
Note that conventional auction methods such as VCG mechanism \cite{Nisan07:AGT}, 
although exhibiting nice properties such as incentive-compatibility, are not guaranteed to converge 
to an equilibrium in an exchange economy with uncertainty. Thus, we use the Walras' tatonnement process 
and prove its convergence in our setting. 
\begin{algorithm}
\caption{Walras' Tatonnement Process}
\label{Alg:Auction}
\small
\begin{algorithmic}[1]
\STATE Select the price adjustment factor $\alpha \to 0$.
\STATE Initialize the price of each good, $p_{m} \to 0$, 
       $m \in \mathcal{M}$.
\REPEAT
\STATE {
\begin{itemize}
 \item Auctioneer announces the prices.
 \item Each consumer announces its demand.
 \item Auctioneer observes excess demands.
 \item Auctioneer adjusts the prices using (\ref{eq:AuctionO}) or (\ref{eq:AuctionT}).
 \end{itemize}}
\UNTIL {Market clears}.
\end{algorithmic}
\end{algorithm}
The following theorem describes the convergence characteristics of Walras' 
tatonnement process.
\begin{theorem}[Convergence of Walras' Tatonnement]
\label{Th:Conv}
Consider an economy $\Omega$ and suppose that $\mathbf{P}$ is a Walrasian (or Arrow-Debreu) equilibrium 
price vector. 
\begin{itemize}
\item Suppose that goods are divisible and the aggregate excess demand function $z(\cdot)$ satisfies 
      the gross substitutes condition. Then the tatonnement process with price adjustment rule (\ref{eq:AuctionO}) 
      converges to the relative prices of $\mathbf{P}$ as $t \to \infty$ for any initial condition 
      $\mathbf{P}(0)$ \cite{Levin06:GE}. 
\item Suppose that goods are indivisible and users' preferences satisfy the gross substitutes condition. 
      Then the tatonnement process with price adjustment rule (\ref{eq:AuctionT}) converges to the 
      relative prices of $\mathbf{P}$ in a finite number of steps for initial prices $\mathbf{P}(0) \to 
      \mathbf{0}$ \cite{Gul99:WEGS}, \cite{Mochaourab15:DCA}.
\end{itemize}      
\end{theorem}
\section{Competitive Market Model of Small Cell Networks}
\label{sec:CMModel}
As described shortly in Section \ref{sec:Wahlras}, in this paper, we model 
the small cell network by an exchange economy under uncertainty. The 
joint residual energy level of small cells corresponds to nature's state, 
which varies randomly and is unknown a priori, hence counts as the source 
of uncertainty. Consumers and commodities, on the other hand, represent 
respectively small cells and users. This implies that in our market model 
only one unit of each commodity exists. As described in Section \ref{subsec:Competitive}, 
in an exchange economy, goods can be divisible or indivisible; this generality 
of our model allows us to associate every user to a single SBS, or let it 
be served by multiple SBSs. We assume that the initial user endowment of 
each small cell is determined by the MBS. For the continuous case, we assume 
that initially the MBS assigns to each SBS some positive fraction of each 
user. Although this can be done simply at random, a smarter choice is to 
allocate initial endowments based on the possible amount of harvested energy. 
The reason is as follows. According to the competitive market 
model described in Section \ref{subsec:Competitive}, each SBS has to pay 
a price for each user. Larger initial endowment thus might correspond to 
higher budget, thereby higher purchase ability; that is, it is reasonable 
to allocate larger endowments, thus larger budget, to those small cells 
that might harvest larger amount of energy and \textit{vice versa}. In the 
case of ambient energy harvesting, this could be done, for instance, by using 
weather history and/or forecast. 

At state $s$, the net utility of small cell $n \in \mathcal{N}$ yields:
\begin{equation}
\label{eq:NetUti}
v_{n}^{(s)}=\sum_{m \in \mathcal{M}}u_{nm}^{(s)}x_{nm}^{(s)}-\sum_{m \in \mathcal{M}}
p_{m}^{(s)}x_{nm}^{(s)},
\end{equation}
where $u_{nm}^{(s)}$ is defined based on the transmission model, as described previously 
in Section \ref{sec:System}. Moreover, $x_{nm}^{(s)}$ is the fraction of user $m$ associated 
to small cell $n$ if state $s$ occurs, where $x \in \{0,1\}$ and $x \in \left[0,1\right]$ 
for discrete and continuous cases, respectively. In cases of fixed-power and equal-power 
transmission models (see Section \ref{sec:System}), for every SBS $n$, the utility achieved 
by serving any user $m \in \mathcal{M}$ depends on the set of assigned user, $\mathcal{M}_{n}$, 
which is unknown when the optimization problem (\ref{eq:UserOpt}) is solved by the SBS. In 
order to solve this issue, we modify the defined utility functions, by assuming the following: 
1) In fixed-power transmission, each SBS $n$ transmits to \textit{all} (not only a subset of) 
its assigned users $m \in \mathcal{M}_{n}$ with fixed power $P_{n}$. By using this approximation, 
the optimization problem actually maximizes an upper-bound of the expected utility. 2) In 
equal-power transmission, each SBS $n$ transmits to all its assigned users with power $P_{n}^{(s)}
=P^{(s)}_{n}/M$. By using this approximation, the optimization problem actually maximizes a 
lower-bound of the expected utility. In Section \ref{subsec:SimLarge} and by means of numerical 
evaluations, we show that such approximations perform well. For the full-power transmission 
model no approximation is required.

Based on the model of exchange economy under uncertainty, users are assigned to small cells 
before the nature's state is known, i.e., before the uncertainty is revealed. Thus, every 
small cell has to pay the price of each one of its selected users, regardless of being able 
to appropriately serve that user or not. In other words, even if low (or no) benefit can be 
made from some user at the occurred state (due to lack of energy), the price of that specific 
user has to be paid. By (\ref{eq:NetUti}), in such cases, the net utility of a small cell can 
become even negative. This mechanism prevents the SBSs from overestimating their ability 
to harvest the energy, and guides the system towards optimality and truthfulness. Moreover, this 
implies that the utility score of each individual depends on the energy level, or in other words, 
the state of nature, which is a random variable (see Section \ref{sec:System}). In essence, the 
difference between conventional deterministic competitive market models (for an example in the 
context of channel assignment, see \cite{Mochaourab15:DCA}) and that under uncertainty is as follows: 
In the latter, every small cell (consumer) pays a price for every user (unit of commodity) before 
it knows its actual energy level (state of the nature), whereas in the former, all information are 
given a priori, excluding all forms of uncertainty. The following theorem describes the equilibrium's 
characteristics in our setting.
\begin{theorem}
\label{Th:OptModel}
Arrow-Debreu equilibrium exists in our setting and is Pareto optimal. Moreover, 
it is unique for the case where a user can be assigned multiple SBSs.
\end{theorem}
\begin{IEEEproof}
\subsubsection{Existence and Pareto optimality}
\begin{enumerate}[(a)]
\item Divisible goods: The utility function defined by (\ref{eq:UtiSmal}) is the sum of continuous, 
      increasing and concave functions. Therefore Assumption \ref{ass:Uti}-a through \ref{ass:Uti}-c 
      holds. On the other hand, \ref{ass:Uti}-d holds due to our system model described in Section 
      \ref{sec:CMModel}. Hence, by Theorem \ref{Th:Ex}-a, Arrow-Debreu equilibrium exists, which is 
      also Pareto optimal due to Theorem \ref{Th:Opt}-a.
\item Indivisible goods: The utility function defined by (\ref{eq:UtiSmal}) is additive separable, 
      since it is completely specified by the values it assigns to singletons. Then, by Lemma \ref{Lm:GSAC}, 
      it satisfies the gross substitutes condition. As it is also monotone, it satisfies the axioms 
      of Assumption \ref{ass:UtiId}. Then, by Theorem \ref{Th:Ex}-b, Arrow-Debreu equilibrium exists, 
      which is also Pareto optimal due to Theorem \ref{Th:Opt}-b.  
\end{enumerate}
\subsubsection{Uniqueness}
Since the utility function given by (\ref{eq:UtiSmal}) is additive separable, it satisfies the gross 
substitutes property by Lemma \ref{Lm:GSAC}. Then, by Lemma \ref{Lm:ExcDem}, the excess demand function 
also satisfies the gross substitutes property. The result therefore follows by Theorem \ref{Th:Uniq}. 
\end{IEEEproof}
The Walras' tatonnement process described in Section \ref{subsec:Auction} can be 
used to implement equilibrium. The entire user association procedure is summarized 
in Algorithm \ref{Alg:UsrAsg}.
\begin{algorithm}
\caption{User Selection with Exchange Economy Model under Uncertainty}
\label{Alg:UsrAsg}
\small
\begin{algorithmic}[1]
\STATE At $t=0$, when the amount of harvested energy at each node is unknown:
       \begin{itemize}
        \item The MBS assigns each SBS an initial (probably state-dependent) endowment, 
              either using available information or at random or simply equally (deterministic).
        \item Each SBS assigns a probability distribution to the set of states. If no 
              information is available, select a distribution uniformly randomly from 
              the $S$-dimensional probability space. 
        \item Perform the Walras' tatonnement process described in Algorithm \ref{Alg:Auction}.
        \item Make commitments.
       \end{itemize} 
\STATE At $t=1$, the amount of available energy in each small cell, i.e., the natures' state 
       is revealed. Each SBS serves the users it has bought under the occurred state $s$, and 
       pays the price. 
\end{algorithmic}
\end{algorithm}
\subsection{Characteristics of Walras' Tatonnement Process in Our Setting}
\label{subsec:Cha}
Primarily, the Walras' tatonnement process necessitates the existence of a coordinator, 
also called auctioneer; nevertheless, it can be implemented in a distributed manner, as 
it is asynchronous and decentralized with respect to both agents and markets \cite{Cheng98:TWA}. 
That is, the price adjustment process is not required to be simultaneous for all commodities 
or for all states. In a small cell network, the MBS can act as an auctioneer, since the 
process imposes low overhead and computational cost, and requires no information except 
for bids. However, as mentioned above, in case an MBS is not available, the Walras' 
tatonnement process can be performed also in a distributed manner, in which SBSs exchange 
their demands with each other. Each SBS then updates the price and announces its new demand 
to others, until market clears. The following theorem describes the convergence characteristics 
of Walras' process in our model.
\begin{theorem}
\label{Th:Converg}
Let $\left(\mathbf{X},\mathbf{P} \right)$ be Arrow-Debreu equilibrium. Then, in our setting,  
the Walras' tatonnement process with price adjustment rule (\ref{eq:AuctionO}) or (\ref{eq:AuctionT}) 
converges to $\left(\mathbf{X},\mathbf{P} \right)$.
\end{theorem}
\begin{IEEEproof}
As described in the proof of Theorem \ref{Th:OptModel}, the utility function given 
by (\ref{eq:UtiSmal}) satisfies the gross substitutes condition for both divisible 
and indivisible goods. Then by Lemma \ref{Lm:ExcDem}, the excess demand function 
satisfies the gross substitutes condition as well. Therefore, the result follows 
by Theorems \ref{Th:Conv}-a and \ref{Th:Conv}-b respectively for divisible and 
indivisible goods. 
\end{IEEEproof}
The convergence speed of Walrasian auction cannot be determined rigorously. In 
essence, the price adjustment path is the main determinant of the speed of convergence, which 
itself depends on excess demand function ($z \left(\mathbf{P}\right)$) as well as the price 
increment factor $\alpha$. On the one hand, larger $\alpha$ yields larger adjustment steps and 
increases the speed of convergence; on the other hand, $\alpha$ being too large prevents the 
convergence, since some commodities would not be interesting to any users if the price becomes 
suddenly too large, and the market does not clear.

The complexity of Arrow-Debreu equilibrium is linear in the number of agents $N$, 
since excess demand function should be built up based on individual demands. Moreover, calculating 
such equilibrium requires to solve a fixed point equation that yields a complexity exponential in 
the number of state contingent commodities, i.e., $SM$ \cite{Axtell05:TEE}.
\subsection{Demand Calculation as a Static Knapsack Problem}
\label{subsec:KSDemand}
Now we will describe how the demand of a  consumer can be modeled and calculated as 
a static knapsack problem. As described in Section \ref{subsec:Competitive} and by 
(\ref{eq:UserOpt}), each consumer has to calculate its demand given prices. For 
divisible goods, where the utility function is concave, the demand can be calculated 
efficiently using conventional convex analysis methods. For indivisible goods, however, 
the problem is more challenging due to its combinatorial nature. Hence, in what follows, 
we provide a model to efficiently calculate the consumers' demands in an economy with 
indivisible goods. 

We model the demand calculation as a \textit{knapsack problem}. The knapsack problem 
is an instance of combinatorial optimization, stated as follows: Given a set of items, 
each item with a size and a value, desired is to determine the number of each item to 
include in a collection so that the total size of selected items does not exceed a 
pre-determined limit (capacity), while the total value is maximized. Let $\mathcal{M}$ 
be the set of items. For each item $m \in \mathcal{M}$, $q_{m}$ is the number of available 
copies, and $x_{m}$ denotes the quantity of item $m$ to be included in the collection. The 
value and size of each item $m \in \mathcal{M}$ are, respectively, shown by $u_{m}$ and 
$p_{m}$. Also, let $B$ stand for the limit (capacity). The problem can be stated formally 
as follows:
\begin{equation}
\label{eq:KSproblem}
\begin{matrix}
\textup{maximize}& \sum_{m\in \mathcal{M}}u_{m}x_{m} \\ 
\textup{s.t.}  &\sum_{m \in \mathcal{M}}p_{m}x_{m} \leq B
\end{matrix}.
\end{equation}
In \textit{0-1 knapsack problem}, there exists only one copy of each item. That is, either 
an object is included in the selected collection or not ($x_{m} \in \{0,1\}$). 

In an exchange economy with indivisible goods, for each state $s \in \mathcal{S}$ and each 
consumer $n \in \mathcal{N}$, calculating the demand of commodities can be modeled as a 
knapsack problem, where initial endowment and budget represent number of copies and limit 
(capacity), respectively. Value and size, on the other hand, stand for utility and price, 
correspondingly. Note that, if there exists only one unit of each commodity, i.e., $y_{m} 
\in \left \{0,1 \right\}$ (as in our setting), then the problem boils down to 0-1 knapsack 
problem. The knapsack problem has been under investigation for over a century, and a variety 
of efficient approximation solutions have been developed. Examples can be found in \cite{Martel90:KPA}. 
Thus, using this model, efficient algorithmic approximate solutions can be used by each consumer 
to calculate her demand. 
\section{Numerical Results and Discussions}
\label{sec:Num}
In this section, in two separate parts, we investigate the proposed model and solution numerically. 
First, we discuss some toy examples in general exchange economy, with our goal being to clarify the 
theoretical model and the solution. Afterwards, we perform extensive simulations in a wireless network, 
in order to evaluate the performance of the proposed model.
\subsection{Toy Examples}
\label{subsec:SimExam}
%
\subsubsection{Example One}
\label{subsec:ExaNState}
\begin{figure}[t]
\centering
\includegraphics[width=0.45\textwidth]{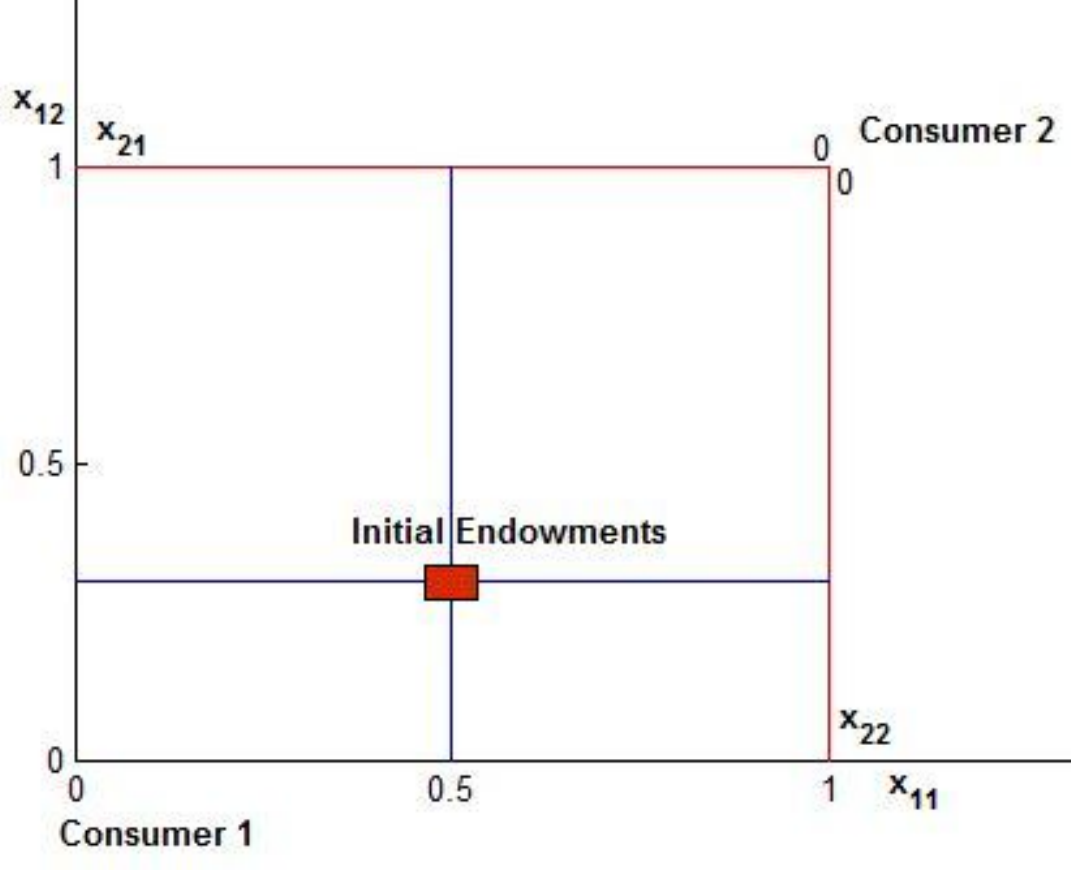}
\caption{Edgeworth box, Example one.}
\label{Fig:EdgB}
\end{figure}

\begin{figure}[t]
\centering
\includegraphics[width=0.45\textwidth]{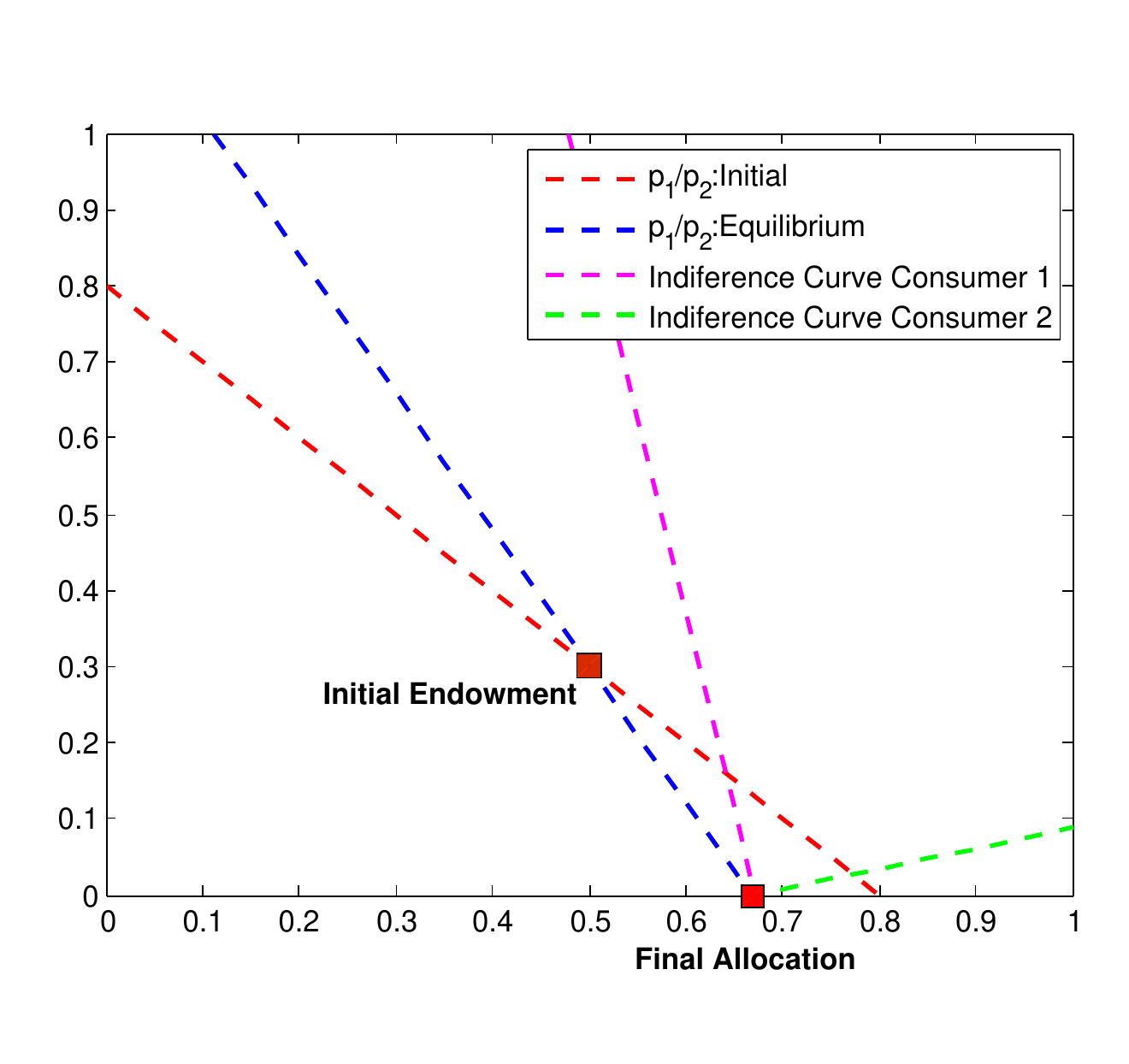}
\caption{Convergence to equilibrium, Example one.}
\label{Fig:EdgC}
\end{figure}

In order to describe the exchange economy model with an example, we consider a 
competitive market with two divisible goods ($M=2$),\footnote{Throughout this 
section, we assume that one unit of each commodity exists.}~two consumers ($N=2$), 
and one state ($S=1$). A single-state scenario implies the absence of uncertainty.\footnote{This 
model is considered here so that important concepts, such as equilibrium, can 
be visualized in two dimensions.}~For the auction process, we let the price 
increasing factor $\alpha=10^{-4}$. The allocation vector is of the form $\mathbf{X}_{i}
=(x_{i1},x_{i2})$, $i \in \{1,2\}$. Let the utility functions $u_{1}$ and $u_{2}$ 
be given as\footnote{Note that here no uncertainty is incorporated, thus we 
use \textit{utility} instead of \textit{expected utility}.}
\begin{equation}
\label{eq:UtiOtr}
u_{1}\left(\mathbf{X}_{1}\right)=0.67x_{11}+0.13x_{12},
\end{equation}
and
\begin{equation}
\label{eq:UtiTtr}
u_{2}\left(\mathbf{X}_{2}\right)=1.62x_{21}+6.01x_{22}.
\end{equation}
For this simple example, the reward sharing can be illustrated by using the \textit{Edgeworth 
box} \cite{Colell85:MT}, shown in Fig. \ref{Fig:EdgB}. The Edgeworth box is a rectangular 
diagram with consumer 1 and consumer 2 origins being located on bottom left and upper right 
corners, respectively. The width and height of the box show the total amount of commodities, 
here one. The bottom line is the $x$-axis for consumer 1 and the left side is the $y$-axis. For 
consumer 2, everything is flipped upside down and backward. Every point in the box represents 
a non-wasteful division of available commodities between the two consumers \cite{Levin06:GE}. 
For instance, from Fig. \ref{Fig:EdgB}, initial endowments read $(x_{11},x_{12})=(0.5,0.3)$ 
and $(x_{21},x_{22})=(0.5,0.7)$. Given the price vector $\mathbf{P}=(p_{1}, p_{2})$, the budget 
line, i.e., the line with slope $p_{1}/p_{2}$ through the endowment point, divides the 
Edgeworth box into two budget sets $\mathcal{B}_{1}\left(\mathbf{P}\right)$ and $\mathcal{B}_{2}
\left(\mathbf{P}\right)$. Then each consumer selects its demand so as to maximize its utility 
given the budget, by using the well-known indifference curves. When supply equals demand, market 
clears and equilibrium is reached. For the exemplary economy described before and for both initial 
and equilibrium prices, the budget lines are shown in Fig. \ref{Fig:EdgC}. Moreover, by using 
indifference curves, convergence to equilibrium is illustrated. It can be seen that through the 
auction process, prices change; as a result, demands change until the convergence condition, 
i.e., market clearing, is satisfied. At equilibrium, $(x_{11},x_{12})=(0.67,0)$ and $(x_{21},
x_{22})=(0.33,1)$. Note that in the presence of uncertainty, 1) commodities are state contingent 
and 2) \textit{expected} utility functions are maximized; The overall procedure is however similar.
\subsubsection{Example Two}
\label{subsec:DisExa}
\begin{figure}[t]
\centering
\includegraphics[width=0.49\textwidth]{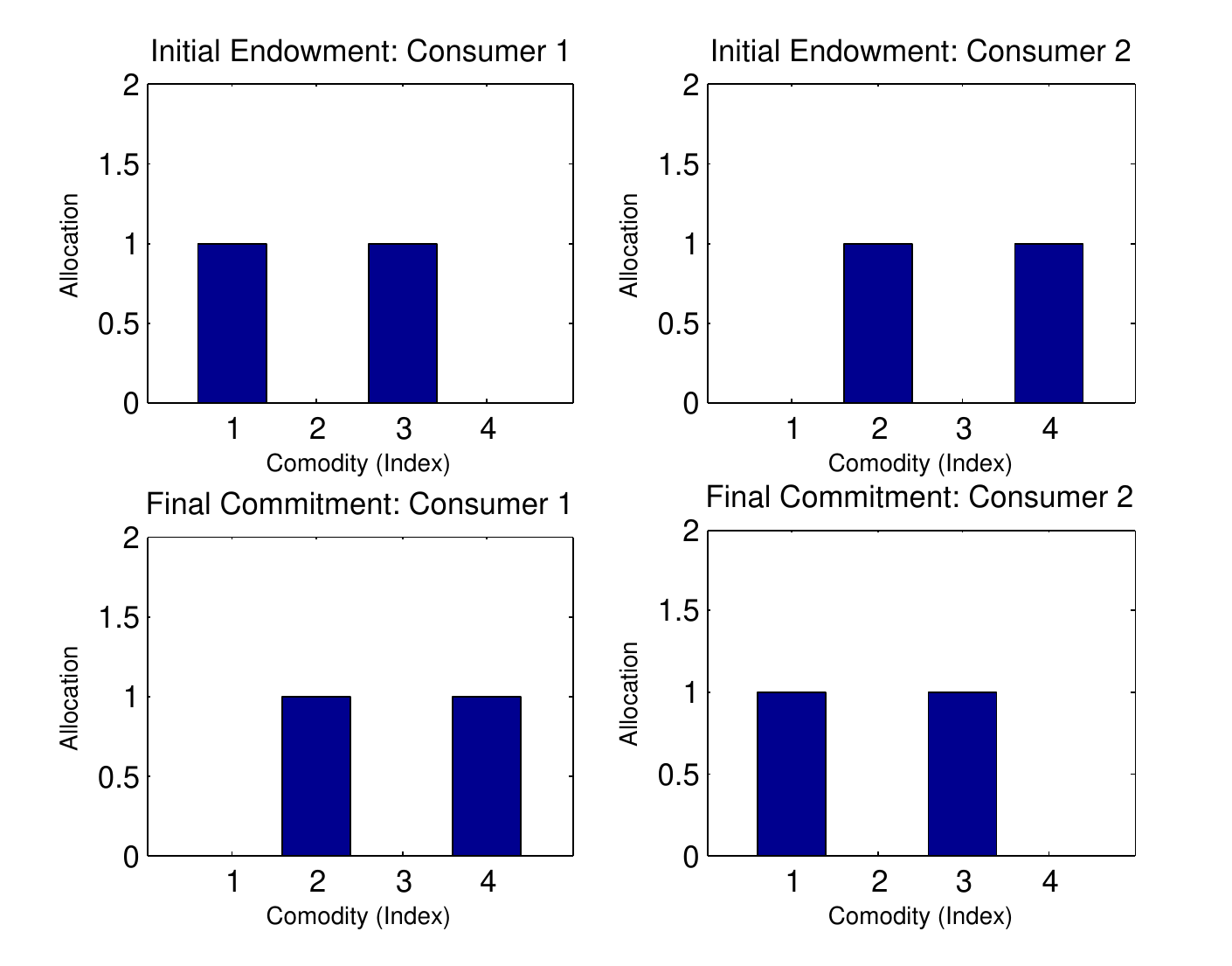}
\caption{Initial endowments and final commitments for Example two (indivisible goods). 
Indices are as follows: $1:x^{(1)}_{i1}, 2:x^{(1)}_{i2}, 3:x^{(2)}_{i1}, 4:x^{(2)}_{i2}$, 
where $i=1$ for consumer 1 and $i=2$ for consumer 2.}
\label{Fig:SimExD}
\end{figure}

In order to clarify the model and solution concept, in this section we consider 
an exchange economy consisting of two indivisible goods ($M=2$) and two consumers 
($N=2$). We assume that the nature has two states ($S=2$). For consumer 1 and 
consumer 2, $\mathbf{a}_{1}=\left(0.2,0.8 \right)$ and $\mathbf{a}_{2}=\left(0.5,0.5 \right)$ 
denote, respectively, the probability vectors assigned to the set of possible states. Also, 
as before, we choose $\alpha=10^{-4}$. For states 1 and 2, utility matrices 
$\left[\mathbf{U}^{(i)} \right]_{N \times M}$, $i \in \{1,2\}$, are given as 
$\mathbf{U}^{(1)}=\begin{bmatrix}
0.10 & 1.75 \\ 
2.20 & 2.20 
\end{bmatrix}$ and $\mathbf{U}^{(2)}=\begin{bmatrix}
0.02 & 0.68 \\ 
0.96 & 0.96 
\end{bmatrix}$. Using $\mathbf{U}^{(i)}$ and $\mathbf{a}_{i}$, the expected utility 
functions $\bar{u}_{i}$, $i \in \{1,2\}$, yield  
\begin{equation}
\label{eq:UtiO}
\bar{u}_{1}\left(\mathbf{X}_{1} \right)=0.02x^{(1)}_{11}+0.35x^{(1)}_{12}+
 0.016x^{(2)}_{11}+0.54x^{(2)}_{12},
\end{equation}
and
\begin{equation}
\label{eq:UtiT}
\bar{u}_{2}\left(\mathbf{X}_{2} \right)=1.10x^{(1)}_{21}+1.10x^{(1)}_{22}+
0.48x^{(2)}_{21}+0.48x^{(2)}_{22}.
\end{equation}
We assume that initial endowments are allocated randomly. Initial endowments and 
final allocations are shown in Figure \ref{Fig:SimExD}. It can be seen that each 
consumer is assigned one good initially. Moreover, from the figure and also by 
(\ref{eq:UtiO}) and (\ref{eq:UtiT}), the Pareto optimal equilibrium is to allocate 
good 1 and good 2 respectively to consumer 2 and consumer 1, regardless of the 
occurred state. Note that in this specific example, contracts are identical for 
all states; this is however not the case in general. Relative prices are here 
$p^{(1)}_{1}/p^{(1)}_{2}=1$ and $p^{(2)}_{1}/p^{(2)}_{2}=1$.   
\subsubsection{Example Three}\label{subsec:ConExa}
\begin{figure}[t]
\centering
\includegraphics[width=0.49\textwidth]{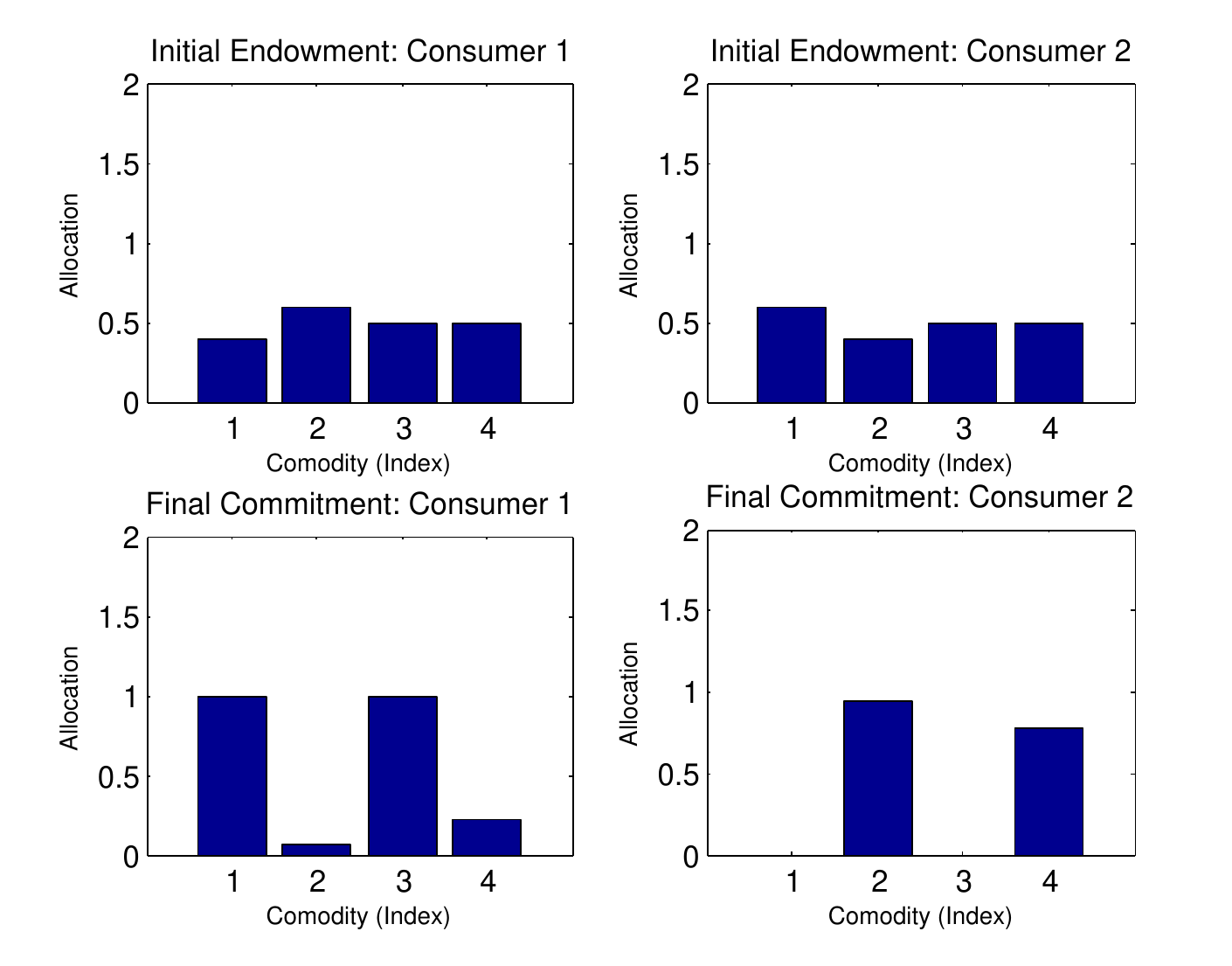}
\caption{Initial endowments and final commitments for Example three (divisible goods). 
Indices are as follows: $1:x^{(1)}_{i1}, 2:x^{(1)}_{i2}, 3:x^{(2)}_{i1}, 4:x^{(2)}_{i2}$, 
where $i=1$ for consumer 1 and $i=2$ for consumer 2.}
\label{Fig:SimExC}
\end{figure}

\begin{figure}[t]
\centering
\includegraphics[width=0.48\textwidth]{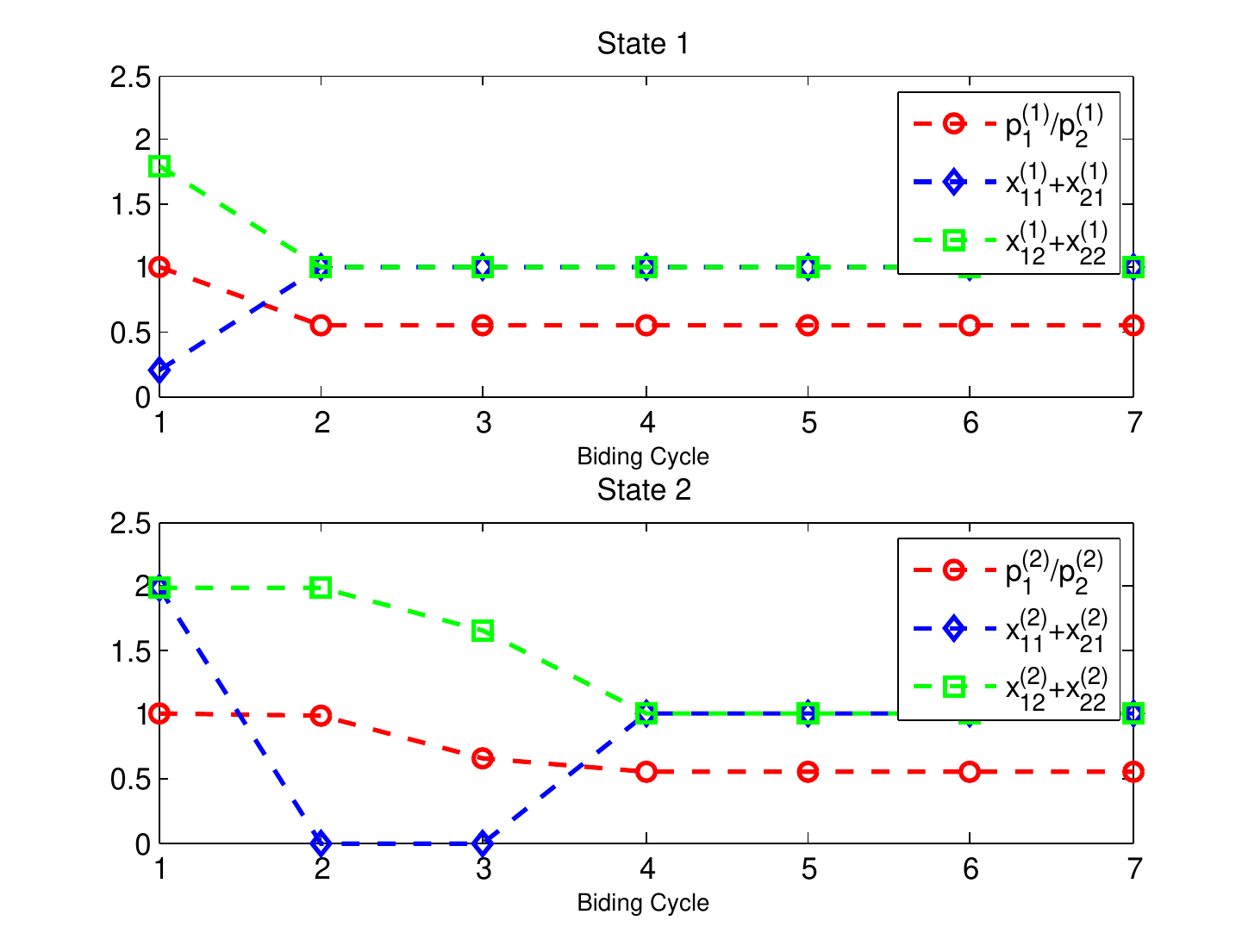}
\caption{Relative prices and aggregate demands, Example three.}
\label{Fig:SimExCT}
\end{figure}

Our third example is similar to the second one, i.e., $M=2$, $N=2$, and $S=2$. Moreover, 
$\alpha=10^{-4}$. In this example, however, we assume that 1) goods are divisible, and 
2) $\mathbf{a}_{1}=\left (0.8,0.2 \right )$ and $\mathbf{a}_{2}=\left (0.5,0.5 \right)$. 
Let $\mathbf{U}^{(1)}=\begin{bmatrix}
1.25 & 2.12 \\ 
0.34 & 2.36 
\end{bmatrix}$ and $\mathbf{U}^{(2)}=\begin{bmatrix}
2.55 & 1.55 \\ 
0.46 & 2.20 
\end{bmatrix}$. Then the expected utility functions $\bar{u}_{1}$ and $\bar{u}_{2}$ yield  
\begin{equation}
\label{eq:UtiOO}
\bar{u}_{1} \left(\mathbf{X}_{1} \right)=x^{(1)}_{11}+1.7x^{(1)}_{12}+0.31x^{(2)}_{11}+0.51x^{(2)}_{12},
\end{equation}
and
\begin{equation}
\label{eq:UtiTT}
\bar{u}_{2} \left(\mathbf{X}_{2} \right)=0.17x^{(1)}_{21}+1.18x^{(1)}_{22}+0.23x^{(2)}_{21}+1.10x^{(2)}_{22}.
\end{equation}
The initial endowments and final allocations are shown in Fig. \ref{Fig:SimExC}.
For both states, relative prices and relative demands of goods are shown in Fig. 
\ref{Fig:SimExCT}. It can be seen that the prices of goods in excess demand increase, 
yielding the demands to decrease. As a result, the demands of other goods either do 
not change or increase, since goods are gross substitutes. The process continues 
until market clears; that is, a price vector is reached in which demand equals supply.
\subsection{Performance Evaluation}
\label{subsec:SimLarge}
In this section, we consider a network with ten users ($M=10$) and four SBSs ($N=4$). 
The average channel gain matrix between SBSs and users, $\left [ \mathbf{H} \right]_{N\times M}$, 
is given as 

$\mathbf{H^{^{\circ\frac{1}{2}}}}=10^{-2}\times \begin{bmatrix}
10 & 71 & 8 & 99 & 99 & 99 & 49 & 99 & 21 & 89\\
11 & 91 & 96 & 95 & 89 & 9 & 21 & 37 & 17 & 30\\
1 & 9 & 92 & 91 & 96 & 11 & 52 & 88 & 52 & 18\\
99 & 19 & 4 & 89 & 92 & 91 & 39 & 7 & 61 & 14
\end{bmatrix}$. 

The components of $\mathbf{H}$ are selected randomly. As 
conventional, we assume that the matrix $\mathbf{H}$ can be written as $\mathbf{H}
=\mathbf{F}{\circ}\mathbf{G}$, where $\mathbf{F}$ and $\mathbf{G}$ are average 
fading gain and path-loss matrices. We let 

$\mathbf{F^{^{\circ\frac{1}{2}}}}=
10^{-2}\times \begin{bmatrix}
10 & 19 & 50 & 21 & 9 & 29 & 4 & 9 & 25 & 8\\
41 & 88 & 32 & 45 & 90 & 9 & 19 & 21 & 8 & 12\\
5 & 6 & 1 & 93 & 98 & 9 & 24 & 40 & 15 & 5\\
50 & 15 & 18 & 92 & 97 & 89 & 8 & 7 & 1 & 2\end{bmatrix}
$. 

According to our system model, initially we assume that every SBS $i$ knows 
only $\mathbf{H}_{i}$; that is, each SBS has only local statistical CSI. The 
antenna gain is initially $G=3$. Noise plus interference power is normalized to one. 
We assume that the nature accepts one of the four possible states; that is, $\mathcal{S}=\left\{
s_{1},s_{2},s_{3},s_{4}\right\}$, where $s_{1}=\left\{2,2,2,2\right\}$, $s_{2}=\left\{6,2,1,5\right\}$,                  
$s_{3}=\left\{3,2,4,7\right\}$, and $s_{4}=\left\{1,3,2,5\right\}$. In $s_{i}$, $i \in \left\{1,
2,3,4\right\}$, the $j$-th component is the energy level of SBS $j$ at state $i$. The probability 
distributions assigned to nature's states by SBSs are as follows: $\mathbf{a}_{1}=\left(0.10,0.50,
0.20,0.20 \right)$, $\mathbf{a}_{2}=\left(0.25,0.25,0.10,0.40 \right)$, $\mathbf{a}_{3}=\left(0.35,
0.15,0.10,0.40 \right)$, and $\mathbf{a}_{4}=\left(0.20,0.30,0.30,0.20 \right)$. These probability 
distributions are selected uniformly randomly from the 4-dimensional probability space. Moreover, we 
choose $\alpha=10^{-4}$. It is clear that four sets of energy levels (nature's states) and ten users 
(commodities) result in 40 state contingent commodities. Each SBS (consumer) determines its demand 
so as to maximize its expected average utility, as discussed in Section \ref{sec:Wahlras} and also 
by simple examples in Section \ref{subsec:DisExa}. For evaluations, we consider the following scenarios:
\begin{itemize}
\item Maximum Received Power Assignment (MRP): In this scenario, user association is performed by a 
central unit given average channel gain matrix, $\mathbf{H}$. By means of exhaustive search, every 
user is assigned to the SBS to which it has the maximum average channel gain. Assignment based on 
received power has been widely used to solve the user association problem (e.g., in \cite{Lin15:OUA}).
\item Nearest SBS Assignment (NSBS): In this scenario, user association is performed by a central 
unit given geographical location of users and SBSs, as well as the path-loss exponent. In our model, 
the path-loss matrix $\mathbf{G}$ follows by element-wise division of $\mathbf{H}$ into $\mathbf{F}$. 
We assume that the path-loss exponent is equal for all links; thus larger distance yields larger path 
loss and \textit{vice versa}. By means of exhaustive search, every user is assigned to the SBS to 
which it has the minimum path-loss. It is clear that the performance of MRP method serves as an 
upper-bound for that of NSBS. Distance-based assignment is a conventional method to solve different 
types of association problems (e.g., in \cite{Son11:BSO}).
\item Maximum Weighted Matching Assignment (MWM): Given the average channel matrix $\mathbf{H}$, the 
user association problem is cast as a maximum-weighted matching problem. More precisely, we construct a bipartite 
graph, with one party being the users and the other party being the SBSs. The weight of each edge 
connecting an SBS to a user is the average channel gain. The \textit{Hungarian algorithm} \cite{Kuhn55:THM} 
is used to assign users to SBSs in a way that each user is associated to only one SBS. Matching 
theory has been widely used to solve the resource allocation as well as association problems (e.g., in 
\cite{Maghsudi12:JPA}).
\item Auction Assignment (AUC): In this scenario, our proposed model and algorithm is used for 
distributed user association under uncertainty.
\item Random Assignment (RND): Users are associated randomly.
\end{itemize}
It should be also mentioned that many user association methods cannot be directly compared to 
each other. This is because, as discussed in Section \ref{sec:Introduction} and also summarized 
in Table \ref{Tb:Comp}, every method is designed for a specific system model and aims at optimizing 
a particular performance metric.

In the next step, the MRP, AUC, and RND association methods are used in conjunction with the 
three transmission scenarios described in Section \ref{sec:System}. 

\begin{figure*}[t]
\centering
\includegraphics[width=1.00\textwidth]{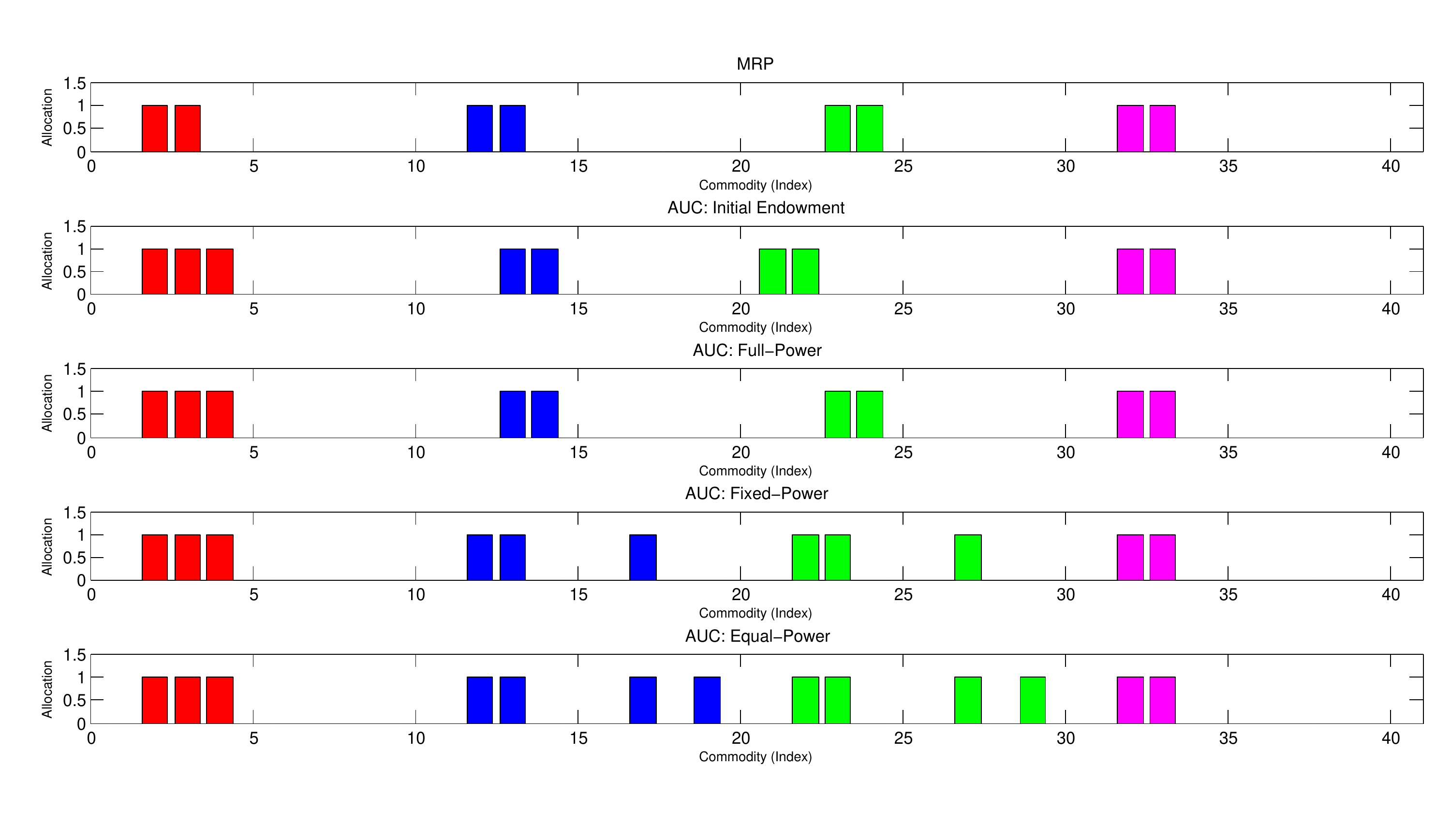}
\caption{Initial endowments and final commitments. Indices $1,2,...,39,40$ correspond to $x_{i1}^{(1)},
x_{i2}^{(1)},...,x_{i9}^{(4)},x_{i10}^{(4)}$, with $i=1$ representing SBS 1.}
\label{Fig:IniFinL}
\end{figure*}

For SBS 1, the allocation 
by MRP, as well as initial endowments and allocations by using the auction process for different 
transmission models are shown in Fig. \ref{Fig:IniFinL}. Note that the MRP allocation is based 
on average channel gain only, hence does not depend on the state and/or transmission model. The 
results are similar for other SBSs. 

\begin{figure*}[t]
\centering
\includegraphics[width=0.99\textwidth]{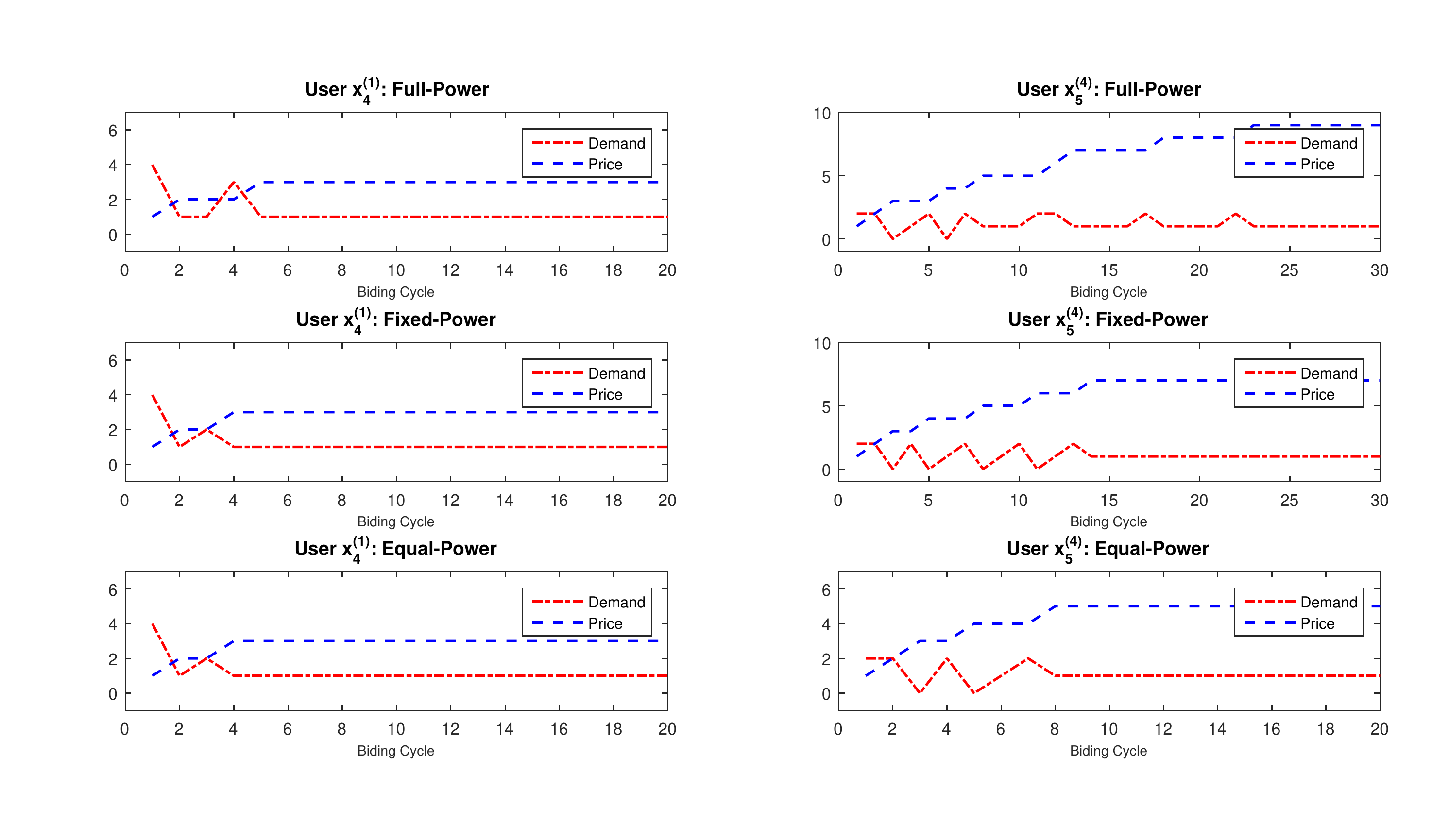}
\caption{Variations in prices and aggregate demands for two exemplary state contingent commodities.}
\label{Fig:PrDe}
\end{figure*}

For AUC and for two exemplary state contingent commodities (User 4 at State 1, $x_{4}^{(1)}$, and 
User 5 at State 4, $x_{5}^{(4)}$), variations in prices and aggregate demands are depicted in Fig. 
\ref{Fig:PrDe}, as a function of auctions' iterations (biding cycles), for different transmission 
models. As expected, excess demand results in price growth. The curves for the rest of state contingent 
commodities are similar.

\begin{figure*}[t]
\centering
\includegraphics[width=0.90\textwidth]{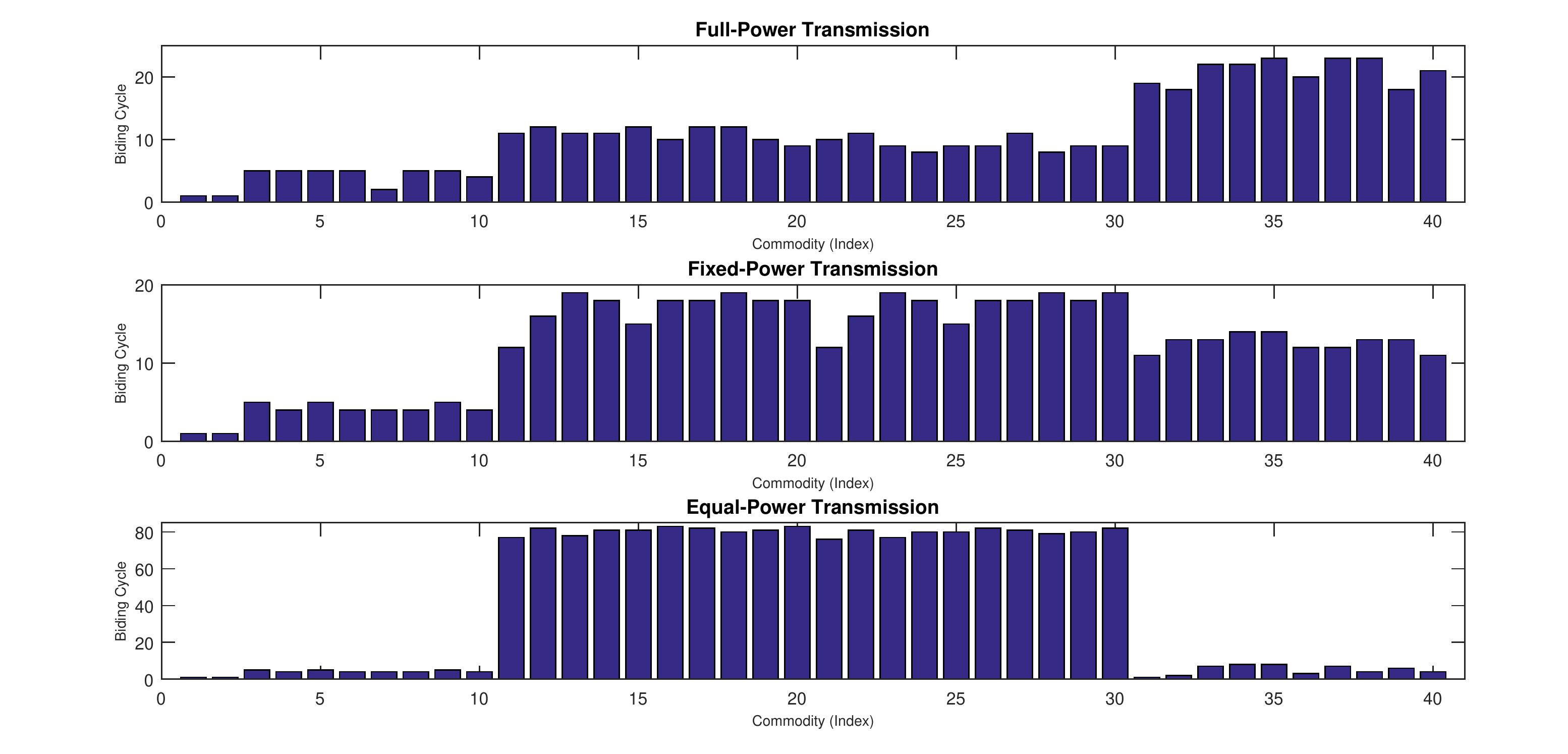}
\caption{Required number of biding cycles to converge for each state contingent commodity. Indices 
$1,2,...,39,40$ correspond to $x_{1}^{(1)},x_{2}^{(1)},...,x_{9}^{(4)},x_{10}^{(4)}$.}
\label{Fig:Speed}
\end{figure*}

The required number of iterations for the price and demand of each good to converge, or, in other 
words, the required number of biding cycles for three (out of four) SBSs to leave the auction, is 
depicted in Fig. \ref{Fig:Speed}. As discussed in Section \ref{subsec:Cha}, the convergence time 
depends on initial endowments and utilities, and cannot be described by a general formula.

\begin{figure*}[t]
\centering
\includegraphics[width=0.93\textwidth]{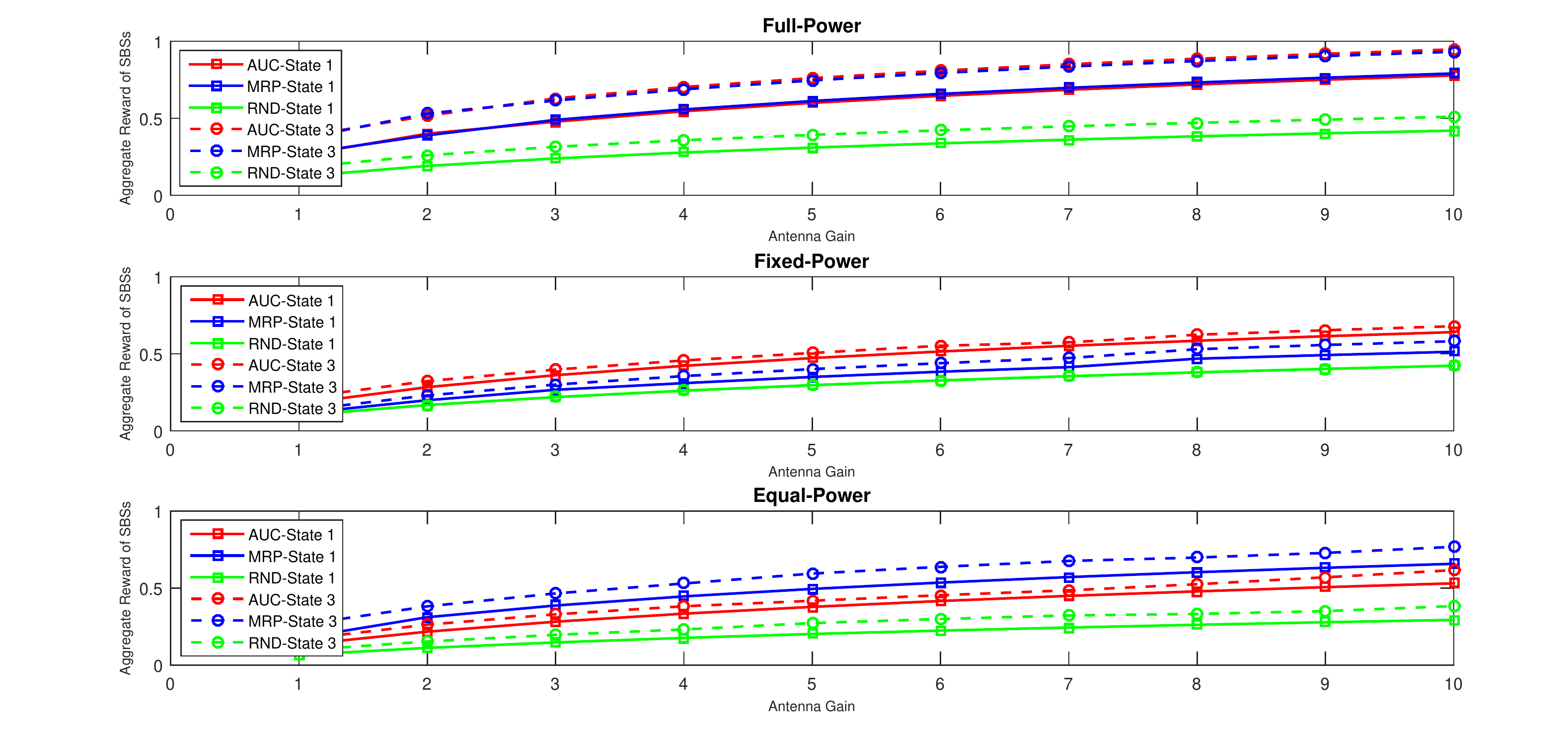}
\caption{Performance of auction model under uncertainty. Aggregate rewards are normalized.}
\label{Fig:Comparison}
\end{figure*}

The performance of the proposed model and solution in terms of aggregate network utility is 
illustrated in Fig. \ref{Fig:Comparison}, for State 1 and State 3. The curves for the other two 
states follow similar paths. 

\begin{figure}[t]
\centering
\includegraphics[width=0.49\textwidth]{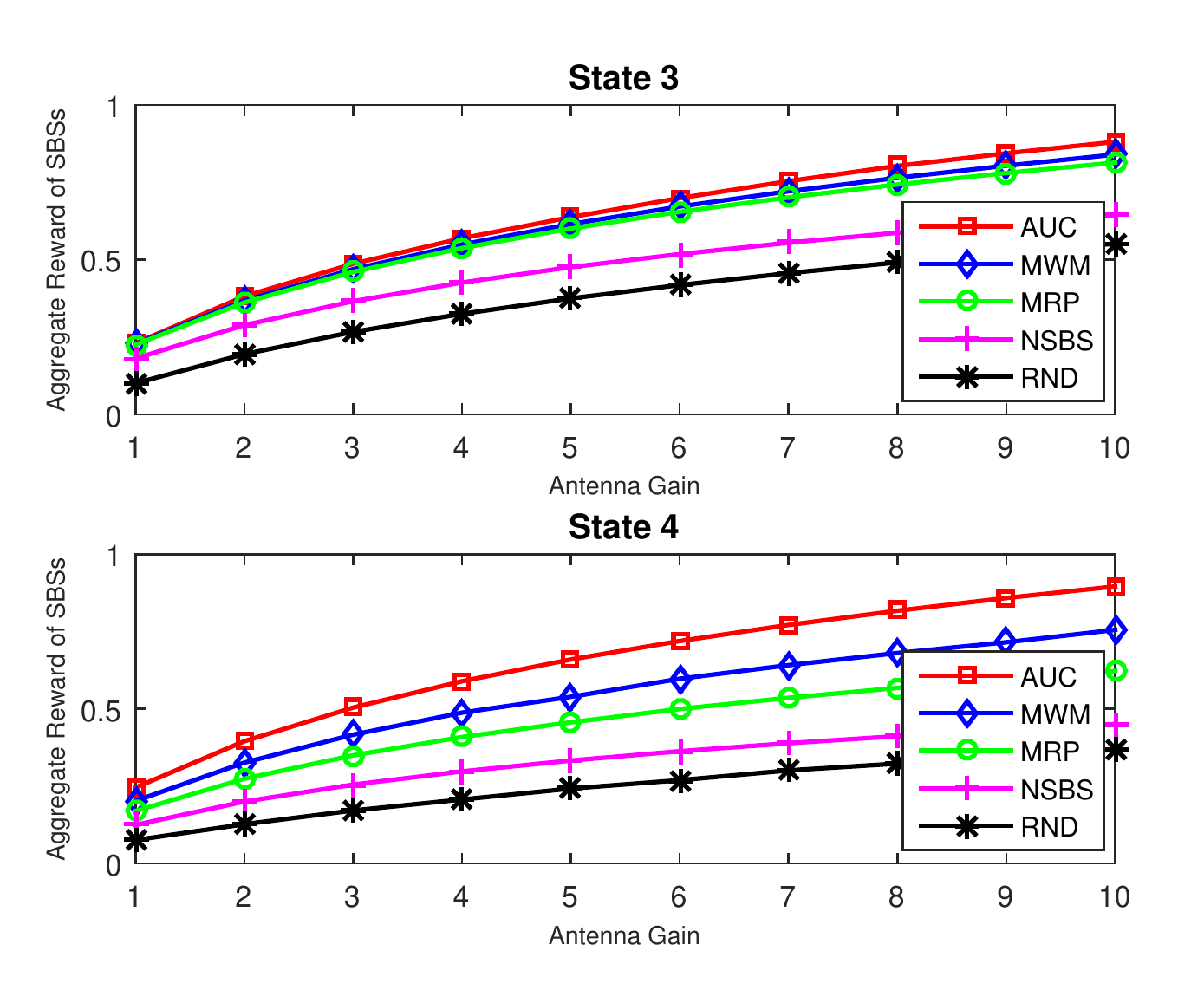}
\caption{Performance of the proposed approach compared to some other user association methods. 
         Aggregate rewards are normalized.}
\label{Fig:Comparisontwo}
\end{figure}

Finally, in Fig. \ref{Fig:Comparisontwo}, we show the performance of the AUC assignment approach 
in comparison with other methods described before, i.e., MRP, NSBS, MWM and RND. As exemplary 
states, we consider State 3 and State 4 in conjunction with fixed-power transmission scenario.   

From Fig. \ref{Fig:Comparison} and Fig. \ref{Fig:Comparisontwo}, it can be concluded that the 
auction model performs well. Moreover, as expected, the performance gain of the auction model 
depends on the transmission model and also varies in different state. The best performance 
appears in fixed-power transmission, where some users might not be served as the result of lack 
of energy. Note that in any case the approach requires little information, yield equilibrium, 
and is implementable in a distributed manner. The occasional small inefficiency is mainly due to 
the uncertainty, as described in the following:
\begin{itemize}
\item In the auction process, SBSs are assigned initial endowments, thereby a budget. As a 
      result, they can select only a specific number of users, depending on the initial 
      endowments. Initial endowments, however, are granted simply at random, leading to an 
      inefficiency in using the harvested energy. Nonetheless, if some information is available 
      at the MBS, for instance some sample data from the past, initial endowments can be granted 
      in a more efficient way, so that an SBS with (possibly) larger harvested energy receives 
      larger budget. Using this policy, the performance of the auction process would be dramatically 
      improved.
\item In the auction process, optimization is performed based on expected utility functions, 
      which, by (\ref{eq:AvNetUt}), depends on probability distributions assigned to the set 
      of states by the SBSs. In the auction model, in general, these distributions are considered 
      to be simply the uniform distribution. However, similar to the previous argument, this 
      assumption yields inefficiency, since it might yield the expected utility functions to 
      be different from true ones.  Therefore, a smarter choice is to select a distribution 
      based on past sample data, and/or by using some forecasting procedure to reduce the 
      effect of uncertainty, so that the expected utility functions are similar to true ones.
\end{itemize}    
%
\section{Conclusion}
\label{sec:con}
We have considered the user association problem in small cell networks, where each 
small cell obtains its energy by using its local energy harvesting units. Desired 
is to develop a distributed user association scheme that is able to cope with 
the uncertainty hidden in the problem, caused by the opportunistic nature of 
energy harvesting. We have modeled the small cell network as a competitive market, 
and the user association as an exchange economy with uncertainty. In our setting, 
we established the existence of Arrow-Debreu equilibrium, and investigated its 
characteristics such as uniqueness and optimality. We have used the Walras' tatonnement 
procedure, combined with the static knapsack problem, to implement equilibrium 
prices efficiently. 
\bibliographystyle{IEEEbib}
\bibliography{references}
\end{document}